\documentclass[aps,prd,superscriptaddress,twocolumn]{revtex4-2}
\usepackage[utf8]{inputenc}
\usepackage{dsfont}
\usepackage{amsfonts}
\usepackage{amsmath}
\allowdisplaybreaks[4]        
\usepackage{amssymb}
\usepackage{euscript}     
\usepackage{braket}
\usepackage{starfont}
\usepackage{color,soul}         
\usepackage{tensor}        
\usepackage{amsthm}
\usepackage{graphicx}
\usepackage{slashed}
\usepackage{leftidx}
\usepackage{subfigure}
\usepackage{bbm}
\definecolor{outerspace}{rgb}{0.25, 0.29, 0.3}
\definecolor{scarlet}{rgb}{1.0, 0.13, 0.0}
\usepackage[header,title,page,titletoc]{appendix}  
\definecolor{princetonorange}{rgb}{1.0, 0.56, 0.0}
\definecolor{WildStrawberry}{rgb}{1.0, 0.26, 0.64}
\definecolor{rossocorsa}{rgb}{0.83, 0.0, 0.0}
\definecolor{navyblue}{rgb}{0.0, 0.0, 0.5}
\usepackage{float}
\usepackage[paper=letterpaper,margin=1in]{geometry}
\usepackage{mathtools}

\DeclareMathAlphabet{\pazocal}{OMS}{zplm}{m}{n}
\newcommand{\req}[1]{(\ref{#1})} 

\newcommand{\bea}{\begin{eqnarray}}

\newcommand{\eea}{\end{eqnarray}}
\newcommand{\ba}{\begin{eqnarray}}
\newcommand{\ea}{\end{eqnarray}}

\newcommand{\be}{\begin{equation}}
\newcommand{\ee}{\end{equation} }
\newcommand{\beqa}{\begin{eqnarray}}
\newcommand{\eeqa}{\end{eqnarray}}
\newcommand{\beqar}{\begin{eqnarray*}}
\newcommand{\eeqar}{\end{eqnarray*}}

\renewcommand{\req}[1]{(\ref{#1})}

\makeatletter
\newcommand{\dal}{\mathop{\mathpalette\dal@\relax}}
\newcommand{\dal@}[2]{%
  \begingroup
  \sbox\z@{$\m@th#1\square$}%
  \dimen0=\fontdimen8
    \ifx#1\displaystyle\textfont\else
    \ifx#1\textstyle\textfont\else
    \ifx#1\scriptstyle\scriptfont\else
    \scriptscriptfont\fi\fi\fi3
  \makebox[\wd\z@]{%
    \hbox to \ht\z@{%
      \vrule width \dimen0
      \kern-\dimen0
      \vbox to \ht\z@{
        \hrule height \dimen0 width \ht\z@
        \vss
        \hrule height 2\dimen0
      }%
      \kern-2.5\dimen0
      \vrule width 2.5\dimen0
    }%
  }%
  \endgroup
}
\makeatother

\usepackage{hyperref}
\hypersetup{
    colorlinks,
    citecolor=navyblue,
    filecolor=navyblue,
    linkcolor=navyblue,
    urlcolor=navyblue
}

\begin{document}

\title{Isospectrality in Effective Field Theory Extensions of General Relativity}

\author{Pablo A. Cano}
\email{pablo.cano@icc.ub.edu}
\affiliation{Departament de F\'isica Qu\`antica i Astrof\'isica, Institut de Ci\`encies del Cosmos\\
 Universitat de Barcelona, Mart\'i i Franqu\`es 1, E-08028 Barcelona, Spain }
 
 \author{Marina David}
 \email{marina.david@kuleuven.be}
 \affiliation{Instituut voor Theoretische Fysica, KU Leuven.
	Celestijnenlaan 200D, B-3001 Leuven, Belgium \vspace{0.1cm}}


\begin{abstract}
Two universal predictions of general relativity are the propagation of gravitational waves of large momentum along null geodesics and the isospectrality of quasinormal modes in many families of black holes. In extensions of general relativity, these properties are typically lost: quasinormal modes are no longer isospectral and gravitational wave propagation is no longer geodesic and it exhibits birefringence --- polarization-dependent speed. We study these effects in an effective-field-theory extension of general relativity with up to eight-derivative terms and show that there is a unique Lagrangian that has a non-birefringent dispersion relation for gravitational waves and isospectral quasinormal modes in the eikonal limit. We argue that both properties are related through a generalized correspondence between eikonal quasinormal modes and unstable photon-sphere orbits. These properties  define a special class of theories that we denote as isospectral effective field theories.
We note that the lowest-order isospectral correction to general relativity coincides with the quartic-curvature correction arising from type II string theory, suggesting that isospectrality might be a key feature of quantum gravity. 
\end{abstract}
\maketitle

General Relativity (GR) represents our current standard model of gravity, as it has passed a plethora of experimental tests and provides a robust dynamical foundation for the gravitational interaction. However, from the point of view of fundamental physics, we know that it is not the end of the story, as it is not an ultra-violet complete theory.  It is now well established that we must approach GR as an effective field theory (EFT) in which the Einstein-Hilbert action is just the leading term at low energies.  As we probe higher energy scales  we are required to include additional terms in the effective action in the form of higher-derivative corrections \cite{Gross:1986iv,Gross:1986mw,Bergshoeff:1989de}.

The details of the EFT depend on the underlying UV-complete theory, so a top-down derivation of the corrections to GR is rarely accessible or practical.  A fruitful alternative to this problem is to consider a bottom-up approach, by writing down a general EFT with all possible higher-curvature invariants up to a specific order \cite{Endlich:2017tqa,Cano:2019ore,Ruhdorfer:2019qmk}. Then, this EFT can be constrained by demanding that it satisfies a set of physically reasonable conditions.  For example, causality conditions (\textit{e.g.} absence of faster-than-light propagation) may constrain the sign and magnitude of the coupling constants of the higher-derivative terms \cite{Adams:2006sv,Gruzinov:2006ie,Brigante:2008gz,Camanho:2014apa,Goon:2016une,deRham:2020ejn,deRham:2020zyh,AccettulliHuber:2020dal, deRham:2021bll, Edelstein:2021jyu, Caron-Huot:2022ugt,Cremonini:2023epg}.
The idea of constraining EFTs based on low-energy conditions is also the basis of the swampland program in string theory \cite{Vafa:2005ui}.

In this work, we propose a new criterion to constrain EFT extensions of GR. Our motivation stems from a feature that appears when studying the propagation of gravitational waves (GW) on curved backgrounds. 

In GR, gravitational waves of large momentum move along null geodesics. This is nothing but a consequence of the Einstein equations under the geometric optics approximation. Furthermore, this statement is independent of the polarization of the wave. 
In extensions of GR, these properties are lost. GWs no longer exhibit geodesic motion as higher-derivative corrections modify their speed of propagation. More strikingly, this speed is different for each polarization mode, as we review below. In other words, the GW propagation becomes \emph{birefringent}. 
This effect depends on the local value of the curvature and hence can be considered as a form of violation of the strong equivalence principle.

A similar phenomenon takes place in the case of perturbations of black holes, characterized by quasinormal modes (QNMs) \cite{Berti:2009kk}. These are modes with specific frequencies and damping times that control the relaxation of the black hole towards the stationary state after experiencing a perturbation. In GR, QNMs of black holes come in two families that are analogous to the two polarization modes of GWs (in fact, QNMs are nothing but GWs on top of a black hole background). A remarkable property that is found in many black hole solutions  --- including Schwarzschild and Kerr black holes --- is that the spectrum of QNM frequencies is identical for both families of modes. This phenomenon, which receives the name of \emph{isospectrality}, is again a benchmark of Einstein gravity that is generically broken in extensions of GR \cite{Cardoso:2009pk,Blazquez-Salcedo:2016enn,Cardoso:2018ptl,Bhattacharyya:2018hsj,deRham:2020ejn,Moura:2021nuh,Cano:2021myl,Cano:2023tmv,Li:2023ulk}. 

Thus, we consider the following question: is there an extension of GR that preserves isospectrality of QNMs and/or non-birefringent GW propagation?
We show that the answer is in the affirmative for both properties as long as one restricts to eikonal QNMs --- those with large angular momentum.  Furthermore, we argue that the two properties are connected: non-birefringence implies isospectrality of eikonal modes. This defines a new class of theories that we will denote as isospectral. 

\textbf{EFT extension of GR:}
Our starting point is the EFT extension of GR with higher-curvature corrections. Up to eight derivatives, the action contains two cubic curvature invariants and three quartic ones, and it is given by \cite{Endlich:2017tqa, Cano:2019ore}
\begin{equation}\label{eq:EFT}
\begin{aligned}
S_{\rm EFT}=\frac{1}{16\pi}\int& d^4x\sqrt{|g|}\bigg[R+\lambda_{\rm ev}\mathcal{I}+\lambda_{\rm odd}\tilde{\mathcal{I}}\\
+&\epsilon_{1}\mathcal{C}^2+\epsilon_{2}\tilde{\mathcal{C}}^2+\epsilon_{3}\mathcal{C}\tilde{\mathcal{C}}+\ldots \bigg]\, ,
\end{aligned}
\end{equation}
with higher-curvature densities
\begin{align}\notag
\mathcal{I}&=\tensor{R}{_{\mu\nu }^{\rho\sigma}}\tensor{R}{_{\rho\sigma }^{\delta\gamma }}\tensor{R}{_{\delta\gamma }^{\mu\nu }}\, ,& \tilde{\mathcal{I}}&=\tensor{R}{_{\mu\nu }^{\rho\sigma}}\tensor{R}{_{\rho\sigma }^{\delta\gamma }} \tensor{\tilde R}{_{\delta\gamma }^{\mu\nu }}\, ,\\
\mathcal{C}&=R_{\mu\nu\rho\sigma} R^{\mu\nu\rho\sigma}\, ,&
\tilde{\mathcal{C}}&=R_{\mu\nu\rho\sigma} \tilde{R}^{\mu\nu\rho\sigma}\, , 
\end{align}
and where
\begin{equation}
 \tilde{R}_{\mu\nu\rho\sigma}=\frac{1}{2}\epsilon_{\mu\nu\alpha\beta}\tensor{R}{^{\alpha\beta}_{\rho\sigma}}
 \end{equation}
is the dual Riemann tensor.  The coefficients $\lambda_{\rm ev, odd}$, $\epsilon_{i}$ are dimensionful coupling constants that we keep arbitrary for now, and we are including both odd ($\lambda_{\rm odd}$, $\epsilon_3$) and even parity ($\lambda_{\rm ev}$, $\epsilon_1$, $\epsilon_2$) corrections for the sake of generality.  We remark that no four-derivative terms appear as they are trivial modulo field redefinitions. 

\textbf{Quasinormal modes of static black holes:}
We start by examining the QNMs of static black holes in the theory \req{eq:EFT}. 
 The gravitational perturbations can be decomposed in two families, axial and polar modes, and can be studied through metric perturbations \cite{Cardoso:2018ptl,deRham:2020ejn,Cano:2021myl}. Here we work instead with the Teukolsky formalism, since it allows us to study isospectrality in an explicit way \footnote{In the metric approach, the perturbations are governed by the Regge-Wheeler \cite{regge1957stability} and Zerilli \cite{zerilli1970effective} equations, for axial and polar perturbations respectively. Although these equations share the same QNM spectrum in the case of Schwarschild black holes, isospectrality is not manifest because the equations take a very different form.}.
In this approach, the black hole perturbations are summarized by the Teukolsky master equations \cite{Teukolsky:1972my}, which are second order equations for the curvature perturbations $\Psi_0$ and $\Psi_4$ --- corresponding to two components of the Weyl tensor. 

Upon separation of variables, the Teukolsky equations can be reduced to radial equations from where one can obtain the QNM frequencies. In the case of Schwarzschild black holes in GR, the Teukolsky radial equations read
\begin{equation}\label{Teq}
\Delta^{-s+1}\frac{d}{dr}\left[\Delta^{s+1}\frac{d R_{s}}{dr}\right]+V_{s} R_{s}=0\,,
\end{equation}
where $s=\pm 2$ and $R_{\pm 2}$ are the radial variables of $\Psi_0$ and $\Psi_4$, respectively. In addition, we have
\begin{equation}
\Delta=r(r-2M)\,,
\end{equation}
and $V_s$ is the Teukolsky potential given by 
\begin{equation}
\begin{aligned}
V_s&=\omega^2r^4+2i s  (r-3 M) \omega r^2+\Delta\left(4+s-l(l+1)\right)\, ,
\end{aligned}
\end{equation}  
where $l=2,3,\ldots$ is the angular momentum of the perturbation and $\omega$ the frequency. An important aspect of the Teukolsky equations with $s=+2$ and $s=-2$ is that they are equivalent: they simply describe the same perturbation using different variables \footnote{In fact, each of the equations can be transformed into the other by using a change of variables: the Teukolsky-Starobinsky identities \cite{Fiziev:2009ud}.}.  Thus, it is enough to consider just one of them. The other crucial aspect is that each equation describes at the same time both families of perturbations: axial and polar \cite{Chandrasekhar:1984siy}. This is the manifestation of isospectrality in GR and the reason why the two families of modes have identical spectra --- they satisfy the same equation. 

Let us then consider the case of the higher-derivative theory \req{eq:EFT}.  A generalized Teukolsky equation for theories beyond GR has been introduced in Refs.~\cite{Li:2022pcy,Hussain:2022ins,Cano:2023tmv} and explicitly applied to the EFT \req{eq:EFT} by Refs.~\cite{Cano:2023tmv,Cano:2023jbk}. The resulting radial equations take the same form as in \req{Teq} but with a modified potential
\begin{equation}\label{shiftV}
V_s\to V_s+\delta V_{s}\, .
\end{equation}
The most crucial aspect about the correction to the potential $\delta V_{s}$ is that, unlike the case of GR, it depends on the type of perturbations one is studying. Specifically, it depends explicitly on two parameters, denoted by $q_{s}$, that characterize the polarization of the perturbation. The value of these parameters is different for each of the two families of QNMs, and this leads generically to a loss of isospectrality.

Thus, we ask the question: is there any choice of coupling constants of the EFT \req{eq:EFT} such that the isospectrality of QNMs of static black holes is preserved?

For the few first harmonics $l=2,3\ldots$, the numerical results for the corrections to the fundamental QNM frequencies can be checked in previous literature \cite{Cardoso:2018ptl,deRham:2020ejn,Cano:2021myl}. Using those results it is not hard to convince oneself that there is no single combination of higher-derivative corrections that gives rises to isospectral modes. Therefore, there is no theory with an exactly isospectral spectrum. 

However, instead of considering the lowest-order harmonics, let us study the eikonal modes $l\rightarrow\infty$.  We focus on the $s=-2$ equation, since, as we mentioned, the $s=\pm 2$ equations are equivalent.
For the cubic theories, using the results of \cite{Cano:2023tmv,Cano:2023jbk}, we find that the leading term in the eikonal expansion of $\delta V_{-2}$ grows with $l^2$ and can be expressed as \footnote{
This limit has to be taken with care, as the form of $\delta V_{s}$ is ambiguous due to the possibility of redefining the radial variables $R_{s}$.  For instance, we find that $\delta V_{s}$ can diverge with a very high power of $l$, but the power of $l$ can be reduced by introducing appropriate transformations of the radial variable. We have used this kind of transformations to obtain the expressions \req{cubicdV} and \req{quarticdV}. We also note that the limit involves taking $l\rightarrow\infty$ and $\omega\rightarrow\infty$ with $\omega/l$ fixed. }
\begin{align}\notag
\delta V_{-2}^{(\rm 6)}=&-(\lambda_{\rm ev}-i\lambda_{\rm odd})\frac{360l^2 M (7 M-3 r)\Delta}{q_{-2}r^6}\\
&+\lambda_{\rm ev}l^2\delta V_0\, ,\label{cubicdV}
\end{align}
where $\delta V_{0}$ is a $q_{s}$-independent term  whose exact form is not relevant for the discussion.  As we can see, this expression depends explicitly on the polarization parameter $q_{-2}$, and this is the reason for the breaking of isospectrality. Each family of QNMs has a different value of $q_{-2}$, and hence a different potential and different QNM frequencies.  

For the sake of completeness, let us explain how the values of the polarization $q_{-2}$ are determined. As explained in \cite{Cano:2023tmv,Cano:2023jbk}, this requires comparing this Teukolsky equation for $\Psi_4$ with the one for the conjugate Teukolsky variable $\Psi_{4}^{*}$, and demanding that both have the same spectrum. It turns out that the conjugate radial equation takes the same form except that the correction to the potential is different
\begin{align}\notag
\delta V_{-2}^{*(\rm 6)}=&-(\lambda_{\rm ev}+i\lambda_{\rm odd})q_{-2}\frac{360l^2 M (7 M-3 r)\Delta}{r^6}\\
&+\lambda_{\rm ev}l^2\delta V_0\, .
\end{align}
Thus, the condition that the two potentials become identical $\delta V_{-2}^{*(\rm 6)}=\delta V_{-2}^{(\rm 6)}$ ---and hence produce the same QNM spectrum --- fixes the polarization to two possible values
\begin{equation}
q_{-2}^{\pm}=\pm \sqrt{\frac{\lambda_{\rm ev}-i\lambda_{\rm odd}}{\lambda_{\rm ev}+i\lambda_{\rm odd}}}\, .
\end{equation}
Since the potentials for the $+$ and $-$ polarizations are different, each family of modes has different QNM frequencies, hence breaking isospectrality. This can only be avoided if the term that depends on  $q_{-2}$ in \req{cubicdV} vanishes, which only happens if $\lambda_{\rm ev}=\lambda_{\rm odd}=0$. 
It follows that there is no cubic theory with isospectral eikonal QNMs. 

Let us then take a look at the eight-derivative theories.  In this case, the correction to the potential scales as $l^4$ in the eikonal limit, and it takes the following simple form
\begin{align}\label{quarticdV}
\delta V_{-2}^{(\rm 8)}=&- \frac{576 l^4 M^2 \Delta}{ r^8} \left(\frac{\epsilon_1-\epsilon_2-i\epsilon_3}{q_{-2}}+\epsilon_1+\epsilon_2\right)\, .
\end{align}
This again depends on the polarization parameter $q_{-2}$ which can be determined in the same way as before.
 The only way to achieve an isospectral result is if the term that depends on $q_{-2}$ vanishes. Since the couplings $\epsilon_{i}$ are real numbers, this only happens if
\begin{equation}\label{eq:epsilon}
\epsilon_1=\epsilon_2\, ,\quad \epsilon_3=0\, .
\end{equation}
This selects the combination $\mathcal{C}^2+\tilde{\mathcal{C}}^2$ as the only theory, up to eighth order in derivatives, that preserves the isospectrality of (eikonal) quasinormal modes of static black holes.   

\textbf{Perturbations of extremal rotating black holes:}
In order to test if the previous result extends to other black holes, it is interesting to consider the case of rotating solutions as well. The Teukolsky equations \req{Teq} have in fact been applied to rotating black holes \cite{Cano:2023tmv,Cano:2023jbk}, but the eikonal limit represents a significant challenge since the methods employed in those papers --- an expansion in the angular momentum --- do not work well with large $l$ modes. 

However, things improve if we consider the extremal limit. In this situation, we can focus on the near-horizon region of the black hole, which is much easier to analyze thanks to its enhanced symmetry. Ref.~\cite{Cano:2024bhh} has recently studied the modified Teukolsky equation for the near-horizon extremal geometries in the EFT of GR \req{eq:EFT}, and has found that the near-horizon Teukolsky radial equation reads
\begin{widetext}
\begin{equation}\label{eq:RadialCorrected}
\begin{aligned}
\rho^{-2s}\frac{d}{d\rho}\left[\rho^{2s+2}\frac{dR_{s}}{d\rho}\right]
+\left(\frac{\bar\omega^2}{\rho^2}+\frac{2\bar\omega(m-2 is)}{\rho}+s+\frac{7m^2}{4}-B_{lm}-\delta B_{lm}\right)R_{s}&=0\, ,
\end{aligned}
\end{equation}
\end{widetext}
Here $\bar\omega$ is the frequency in the near-horizon region, $(l,m)$ are the angular harmonic numbers and $B_{lm}$ are the angular separation constants --- eigenvalues of the angular Teukolsky equation for Kerr that we do not show here.  The only effect of the higher-derivative terms is to modify these angular separation constants by adding a correction $\delta B_{lm}$ --- this is analogous to the correction to the potential in \req{shiftV}.

The corrections to the separation constants $\delta B_{lm}$ depend on the polarization parameters $q_{s}$, just like it happens for the correction to the potential $\delta V_{s}$. We then ask if there is any theory for which the correction $\delta B_{lm}$ is the same for both polarization modes, hence preserving isospectrality.  

As in the case of static black holes, we consider the eikonal limit.   Ref.~\cite{Cano:2024bhh} has shown that, to leading order in the eikonal expansion, $\delta B_{lm}$ is given by
 \begin{equation}\label{magicformula}
 \delta B_{lm}=\int_{0}^{\pi} d\theta F(\theta,x_0) f_{lm}(x_0 \cos\theta)\,, 
 \end{equation} 
 where
 \begin{equation}
 F(\theta,x_0)=\frac{1}{2K\left(\frac{x_0^2(-1+x_0^2)}{3+x_0^2}\right)\sqrt{1+\frac{x_0^2(1-x_0^2)}{3+x_0^2} \cos^2\theta}}\, ,
 \end{equation}
 $K(k)$ is the complete elliptic integral of the first kind, $x_0$ is related to the ratio $m/l$ by
 \begin{equation}
  \frac{m^2}{l^2}=\frac{4(1-x_0^2)}{4+x_0^2-x_0^4}\, ,
 \end{equation}
 and $f_{lm}(x)$ is a theory-dependent function obtained from the modified Teukolsky angular equation. 
For the quartic theories in \req{eq:EFT}, this function reads
\begin{equation}
\begin{aligned}
f_{lm}(x)&=\frac{36\left(4 B_{lm}-3 m^2\right)^2}{M^6}\Bigg[-\frac{\epsilon_1+\epsilon_2}{(1+x^2)^4}\\
&+(\epsilon_1-\epsilon_2-i\epsilon_3)\frac{i (x+i)}{q_{-2} (x-i)^9}\Bigg]\, ,
\end{aligned}
\end{equation}
and as we see, it depends explicitly on the polarization parameter $q_{-2}$. However, the term that depends on $q_{-2}$ vanishes for the theory \req{eq:epsilon}, which henceforth has isospectral Teukolsky equations in the near-horizon region. Furthermore, this is again the only theory with this property, including the cubic theories whose expressions for $f_{lm}(x)$ can be checked in \cite{Cano:2024bhh}.

\textbf{Speed of gravitational waves:}
The previous results motivate us to study a different aspect of the $\mathcal{C}^2+\tilde{\mathcal{C}}^2$ theory: GW propagation. 
In GR, there is a correspondence between eikonal QNMs and unstable null geodesics orbiting the photon sphere of the black hole \cite{Cardoso:2008bp,Yang:2012he}. 
This relationship has its origin in the fact that GWs of large momentum follow null geodesics.  
A similar correspondence between QNMs and GW orbits is expected to hold for theories beyond GR \cite{Bryant:2021xdh}, but in this case things become more involved because GWs no longer follow null geodesics in the large momentum limit. 
Thus, the fact that the theory $\mathcal{C}^2+\tilde{\mathcal{C}}^2$ exhibits isospectral eikonal QNMs, might be telling us that there is something special about the GW propagation in this theory. 

In order to test this, we analyze the dispersion relation of GWs in the EFT of GR \req{eq:EFT}. In the case of the quartic theories, the dispersion relation in the large momentum limit was first analyzed in \cite{Gruzinov:2006ie} (see also \cite{Endlich:2017tqa}), and we provide a more detailed derivation in the appendix. We find 
\begin{equation}\label{eq:dispersionrelation}
\begin{aligned} 
	k^2 &= 64\epsilon_1 (S_{\mu\nu}e^{\mu\nu})^2 + 64\epsilon_2 (\tilde{S}_{\mu\nu}e^{\mu\nu})^2\\
	&+ 64\epsilon_3 (\tilde{S}_{\mu\nu}e^{\mu\nu})(S_{\alpha\beta}e^{\alpha\beta}) \,,
\end{aligned}
\end{equation}
where $k_{\mu}$ is the wave vector, $e_{\mu\nu}$ is the polarization tensor and
\begin{align} \label{eq:defofS}
	S_{\mu\nu} &\equiv k^{\rho}k^{\sigma}R_{\mu\rho\sigma\nu}\,,
	\quad
	\tilde{S}_{\mu\nu} \equiv k^{\rho}k^{\sigma}\tilde{R}_{\mu\rho\sigma\nu}\,,
\end{align}
where $R_{\mu\rho\sigma\nu}$ is the background curvature tensor. 
The polarization tensor is symmetric, traceless and transverse to $k_{\mu}$ --- that is, $k^{\mu}e_{\mu\nu}=0$ --- and it is normalized as $e^{\mu\nu}e_{\mu\nu}=1$.
The tensors $S_{\mu\nu}$ and $\tilde{S}_{\mu\nu}$ are also symmetric and transverse to $k_{\mu}$.
In addition, to first order in the higher-derivative corrections, the right-hand-side of \req{eq:dispersionrelation} is evaluated on a Ricci-flat background.  Therefore, $S_{\mu\nu}$ and $\tilde{S}_{\mu\nu}$  are also traceless. 

In the case of GR, Eq.~\req{eq:dispersionrelation} implies that $k^2=0$, meaning that GWs move at the speed of light. With higher-derivative corrections, $k^2\neq 0$, and as we can observe, the dispersion relation depends explicitly on the polarization of the wave: this is the definition of birefringence. As a consequence, GWs of different polarization propagate at different speeds and follow different trajectories.  
However, something remarkable happens in the case of the special theory \req{eq:epsilon}: the dispersion relation becomes polarization-independent. 

In order to prove this we first simplify the second term in \eqref{eq:dispersionrelation} by expanding the product of the two Levi-Civita tensors appearing in each of the dual Riemann tensors. This gives
\begin{align}  \label{eq:Stildeesimp}
	(\tilde{S}_{\mu\nu}e^{\mu\nu})^2 = S_{\nu\lambda}S^{\beta\lambda}e^{\mu\nu}e_{\mu\beta} - S^{\alpha}{}_{\mu}S^{\beta}{}_{\nu}e^{\mu\nu}e_{\alpha\beta}\, .
\end{align}
In addition, there are several identities relating the different contractions of $S_{\mu\nu}$ and $e_{\mu\nu}$. In order to derive them, we consider the following identities arising from the antisymmetrization of five indices
\begin{align} 	
	0&=k^{\rho}k_{[\sigma|}\tensor{S}{_{\mu}^{\nu}}\tensor{R}{_{|\alpha \rho}^{ \sigma \beta}}\tensor{e}{^{\mu}_{\nu}}\tensor{e}{^{\alpha}_{\beta]}}\, ,
	\\
	0&=k^{\rho}k_{[\sigma}\tensor{S}{_{\mu|}^{\nu}}\tensor{R}{_{|\alpha| \rho}^{ \sigma \beta}}\tensor{e}{^{\mu}_{|\nu}}\tensor{e}{^{\alpha}_{\beta]}}\, .
\end{align}
Once expanded and simplified we are left with the following relations
\begin{align} \label{eq:identitiessimp}
	0 =& - (S_{\mu\nu}e^{\mu\nu})^2  + S^{\alpha}{}_{\mu}S^{\beta}{}_{\nu}e^{\mu\nu}e_{\alpha\beta}+  S_{\nu\lambda}S^{\beta\lambda}e^{\mu\nu}e_{\mu\beta} 
	\nonumber \\
	0=& - 2(S_{\mu\nu}e^{\mu\nu})^2  +2S^{\alpha}{}_{\mu}S^{\beta}{}_{\nu}e^{\mu\nu}e_{\alpha\beta}   \nonumber \\&+ 4 S_{\nu\lambda}S^{\beta\lambda}e^{\mu\nu}e_{\mu\beta}  -  S_{\mu\nu}S^{\mu\nu}\, .
\end{align}
With the help of these and \req{eq:Stildeesimp}, the dispersion relation \req{eq:dispersionrelation} for the theory \req{eq:epsilon} reduces to
\begin{align}
	k^2 = 64\epsilon_1 S_{\mu\nu}S^{\mu\nu}\, ,
\end{align}
which is manifestly independent of the polarization. This shows that the theory $\mathcal{C}^2+\tilde{\mathcal{C}}^2$ is non-birefringent, and it is in fact the only quartic theory with this property. In addition, let us take note that this result is general: it happens for gravitational waves propagating on any background solution of the theory. 

For the cubic theories, we expect the dispersion relation to always be birefringent, since eikonal QNMs are not isospectral. We briefly explore this in the appendix.

\textbf{Discussion:}
Up to eight-derivative order, we have identified a unique extension of GR
\begin{equation} \label{eq:Lspecial}
S_{\rm iso}=\frac{1}{16\pi}\int d^{4}x\sqrt{|g|}\left[R+\epsilon_1\left(\mathcal{C}^2+\tilde{\mathcal{C}}^2\right)\right]\,,
\end{equation}
that satisfies very remarkable properties.  This theory (i) gives rise to non-birefringent GW propagation on arbitrary backgrounds and (ii) possesses isospectral eikonal QNMs for (at least) static and extremal rotating black holes. 

The two properties are clearly related: QNMs are nothing but GWs on a black hole background, and the eikonal limit is equivalent to the large momentum limit. Thus, there must be a correspondence between eikonal QNMs and GW orbits around black holes, completely analogous to the one in GR \cite{Cardoso:2008bp,Yang:2012he}.  The fact that large momentum GWs in \req{eq:Lspecial} are non-birefringent means that the two polarization modes follow identical trajectories, and the counterpart of this is that the two families of QNMs are isospectral. Therefore, we expect that the theory \req{eq:Lspecial} actually has isospectral eikonal QNMs for black holes of arbitrary angular momentum, not only for the cases we have studied. In fact, we conjecture that any theory with a non-birefringent dispersion relation for GWs satisfies this property.  Thus, we propose to use the term ``isospectral'' to refer to non-birefringent theories, since this seems to be the only notion of isospectrality that can be extended beyond GR \footnote{The QNMs in the theory \req{eq:Lspecial} are in fact only isospectral in the eikonal limit, as isospectrality is broken for lower $l$ modes.  Nevertheless, by using the numerical values of QNMs available in the literature \cite{Cano:2023jbk}, one can see that the isospectrality breaking in the theory \req{eq:Lspecial} is pretty mild even for the lower $l$ modes.}.

The condition of isospectrality, defined in this way, is extremely powerful in constraining the form of the EFT extension of GR. As we have seen, to the order considered it implies that: (a) the six-derivative corrections vanish,  (b) the theory preserves parity and (c) there is a single quartic curvature correction. Based on this, it would be interesting to try to determine the form of the isospectral extension of GR at even higher orders. 

There is a final surprise, as the isospectral EFT \req{eq:Lspecial} is unexpectedly connected to string theory. In fact, in string theory it is possible to find explicitly some of the corrections to the Einstein-Hilbert action, which appear through an expansion in the parameter $\alpha'$. The leading correction in the ten-dimensional effective action of type II theories is a term of the form  $\frac{\zeta(3)\alpha'^3}{8} R_4$ \cite{Gross:1986iv}, which involves a quartic curvature invariant that can be expressed as $R_4=\left(R_{\mu \nu \rho \sigma}R^{\alpha\nu \rho \beta}+\frac{1}{2}R_{\mu \sigma \nu \rho}R^{\alpha \beta\nu \rho}\right)R\indices{^\mu^\tau_\epsilon_\alpha}R\indices{_\beta_\tau^\epsilon^\sigma}$ \cite{Grisaru:1986vi}. 
When reduced to four dimensions and expressed in the basis of the EFT \req{eq:EFT}, this precisely yields \cite{Gruzinov:2006ie,Cano:2022wwo}
\begin{equation}
R_4=\frac{1}{32}\left( \mathcal{C}^2+\tilde{\mathcal{C}}^2\right)\,.
\end{equation}
Remarkably, the isospectral condition exactly reproduces the prediction of string theory for the leading higher-derivative corrections to Einstein gravity \footnote{Although the string theory effective action contains other fields, they can be consistently truncated at this order in $\alpha'$. In particular, the dilaton gets an $\mathcal{O}(\alpha'^3)$ value due to its coupling to the  $R^4$ term, but it only affects the metric at order $\mathcal{O}(\alpha'^6)$. }. 
It would be very interesting to understand the origin of this connection. One may conjecture that it could be related to supersymmetry, or perhaps to the existence of a gravitational duality \cite{Henneaux:2004jw} --- in fact, string dualities may act on the graviton \cite{Hull:2000rr}.  Arguably, such duality would imply a symmetry between the two degrees of freedom of the graviton, leading to isospectrality. This would make the isospectral extensions of GR gravitational analogues of the Born-Infeld theory of electrodynamics \cite{Born:1934gh}, which is both duality invariant and non-birefringent \cite{Bialynicki-Birula:1984daz,Gibbons:1995cv,Russo:2022qvz}.

If one could establish the connection between isospectrality and string theory in general, then isospectrality could be used to constrain the form of additional stringy corrections, which is a challenging problem --- \textit{e.g.} \cite{Bergshoeff:1989de,Marques:2015vua, Baron:2017dvb,  Hohm:2015doa, Codina:2020kvj, Garousi:2020gio, David:2021jqn, David:2022jcl}.  More importantly, this connection would single out the isospectral theories as preferred quantum gravity EFTs. We leave these questions for future work.

\vspace{0.1cm}
\begin{acknowledgments} 
We would like to thank Nikolay Bobev, Roberto Emparan, Robie Hennigar, Tom\'as Ort\'in and Harvey Reall and Thomas Van Riet for useful comments and conversations. The work of PAC received the support of a fellowship from “la Caixa” Foundation (ID 100010434) with code LCF/BQ/PI23/11970032.  MD is supported in part by the Odysseus grant (G0F9516N Odysseus) as well as from the Postdoctoral Fellows of the Research Foundation - Flanders grant (1235324N).
\end{acknowledgments}

\appendix

\section{The dispersion relation}
In this appendix, we derive the dispersion relation for a general EFT extension of general relativity with Lagrangian
\begin{align}
	S_{\text{EFT}} = \frac{1}{16\pi} \int d^4 x \sqrt{|g|} \left[R + \mathcal{L}(R_{\mu\nu\rho\sigma})\right]\,.
\end{align}
The corresponding equations of motion can be written as
\begin{align}\label{EFE2}
	G_{\mu\nu} = T^{}_{\mu\nu}\,,
\end{align}
and the effective stress energy tensor $T_{\mu\nu}^{}$ reads
\begin{align} \label{eq:stressenergy}
	\begin{split}
	T_{\mu \nu}^{} &=-\tensor{P}{_{(\mu}^{\rho \sigma \gamma}} R_{\nu) \rho \sigma \gamma}+\frac{1}{2} g_{\mu \nu} \mathcal{L}_{} - 2 \nabla^\sigma \nabla^\rho P_{(\mu|\sigma| \nu) \rho}^{}\,,
	\end{split}
\end{align}
which explicitly depends on the Lagrangian and the tensor $P_{\mu\nu\alpha\beta}$ 
\begin{equation}\label{defofP1}
	P_{\mu\nu\alpha\beta}=\frac{\partial \mathcal{L}}{\partial R^{\mu\nu\alpha\beta}}\,.
\end{equation}
In order to find the dispersion relation, one must linearize the equations of motion \req{EFE2} around an arbitrary background by performing $g_{\mu\nu} \to g_{\mu\nu} + h_{\mu\nu}$. The linearization of the Einstein tensor is given by
\begin{align} \label{eq:deltaGmunu}
	\begin{split}
	\delta G_{\mu\nu} &= -\frac{1}{2} \left[\nabla^2 h_{\mu\nu} - 2\nabla^{\rho}\nabla_{(\mu}h_{\nu)\rho} 
	+ \nabla_{\mu}\nabla_{\nu}h  \right. \\& \left. + g_{\mu\nu} \left(\nabla_{\rho}\nabla_{\sigma}h^{\rho\sigma} -\nabla^2 h\right) \right] \,.
	\end{split}
\end{align}
Using the de Donder gauge $\nabla_{\mu}h^{\mu\nu}=0$ and further imposing $h_{\mu\nu}$ to be traceless, $h=0$ -- using residual gauge freedom -- this expression simplifies to
\begin{align}
	\delta G_{\mu\nu} = -\frac{1}{2}\nabla^2 h_{\mu\nu}\,.
\end{align}
The linearization of the higher-derivative part of \req{EFE2} is more complicated, but it simplifies in the large momentum limit. 
Thus, we consider the following ansatz for $h_{\mu\nu}$
\begin{align} \label{eq:hansatz}
	h_{\mu\nu} = e_{\mu\nu} e^{i \Phi},
\end{align}
where $e_{\mu\nu}$ is the polarization tensor and $\Phi$ is a certain function whose gradient yields the momentum of the wave
\begin{equation}
\partial_{\mu}\Phi\equiv k_{\mu}\, .
\end{equation}
In the geometric optics approximation $k_{\mu}\rightarrow\infty$, we assume that the polarization $e_{\mu\nu} $ and the momentum $k_{\mu}$ only vary at scales much larger than the wavelength of the perturbation. Thus, to leading order in the large momentum expansion, these two quantities are constant and furthermore satisfy
\begin{equation}
k^{\mu}e_{\mu\nu}=0\, ,\quad g^{\mu\nu}e_{\mu\nu}=0\, ,
\end{equation}
on account of the gauge conditions imposed on $h_{\mu\nu}$. Furthermore, we can always normalize $e_{\mu\nu}$ such that
\begin{equation}
e_{\mu\nu}e^{\mu\nu}=1\, .
\end{equation}
We now have what we need to linearize the equations of motion \eqref{EFE2}. The right hand side of this equation is the effective stress energy tensor \req{eq:stressenergy} which involves three distinct terms. In the geometric-optics limit --- very small wavelengths relative to the size of the radius of curvature --- the scaling of each term in \eqref{eq:stressenergy} is given by
\begin{align} \label{eq:leadingbehaviort1}
	\delta (P^{}_{\mu\nu\rho\sigma}R^{\mu\nu\rho\sigma}) & \sim R_{\mu\nu\rho\sigma}\delta R^{\mu\nu\rho\sigma} \sim \mathcal{O}(k^2)\,,
	\\ \label{eq:leadingbehaviort2}
	\delta (\mathcal{L}_{}) & \sim R_{\mu\nu\rho\sigma}\delta R^{\mu\nu\rho\sigma} \sim \mathcal{O}(k^2)\,,
	\\ \label{eq:leadingbehaviort3}
	\delta (\nabla^2 P^{}_{\mu\nu\rho\sigma}) &\sim \mathcal{O}(k^4)\,,
\end{align}
where we have used $\nabla_{\rho}h_{\mu\nu} \simeq i k_{\rho}h_{\mu\nu}$ and
\begin{align} \label{eq:linearizedRiemann}
	\begin{split}
		\delta R_{\mu\nu\rho\sigma} & = k_{\nu}k_{[\sigma}h_{\rho]\mu} - k_{\mu}k_{[\sigma}h_{\rho]\nu}\,,
	\end{split}
\end{align}
to leading order in large $k^{\mu}$. The subleading terms \eqref{eq:leadingbehaviort1} and \eqref{eq:leadingbehaviort2} can be neglected as the leading term is of order $\mathcal{O}(k^4)$.

With this in mind, the linearized equation of motion boils down to
\begin{align} \label{eq:dispersion}
	k^2 h^{\mu\nu} = 4 k^{\sigma} k^{\rho} \delta P^{}_{(\mu|\sigma|\nu)\rho} + \mathcal{O}(k^3)\,,
\end{align}
where effectively every covariant derivative acting on $h_{\mu\nu}$ gives us a factor of $ik^{\mu}$. 

\subsection{The dispersion relation for $\mathcal{L}_{(8)}=\mathcal{L}(\mathcal{C},\tilde{\mathcal{C}})$}
Let us consider a theory $\mathcal{L}_{(8)}=\mathcal{L}(\mathcal{C},\tilde{\mathcal{C}})$ composed of combinations of the quadratic curvature invariants $\mathcal{C}$, $\tilde{\mathcal{C}}$ up to arbitrary order.

Explicitly, the tensor $P^{(8)}_{\mu\nu\rho\sigma}$ for this theory is
\begin{align}
	\begin{split}
	P^{(8)}_{\mu\nu\rho\sigma} &= 
	2 \frac{\partial \mathcal{L}_{(8)}}{\partial\mathcal{C}} R_{\mu\nu\rho\sigma} +  2\frac{\partial \mathcal{L}_{(8)}}{\partial \tilde{\mathcal{C}}}\hat{R}_{\mu\nu\rho\sigma},
	\end{split}
\end{align}
with $\hat{R}_{\mu\nu\rho\sigma} \equiv \frac{1}{2}(\tilde{R}_{\mu\nu\rho\sigma} + \tilde{R}_{\rho\sigma\mu\nu})$.  Its variation is
\begin{align} \label{eq:deltaPquartic}
	\begin{split}
		\frac{1}{2}\delta P^{(8)}_{\mu\nu\rho\sigma} 
		&= 
		\frac{\partial \mathcal{L}_{(8)}}{\partial\mathcal{C}} \delta R_{\mu\nu\rho\sigma} + \frac{\partial \mathcal{L}_{(8)}}{\partial \tilde{\mathcal{C}}}\delta \hat{R}_{\mu\nu\rho\sigma}
		\\& +\frac{\partial^2 \mathcal{L}_{(8)}}{\partial\mathcal{C}^2} \delta \mathcal{C} R_{\mu\nu\rho\sigma} + \frac{\partial^2 \mathcal{L}_{(8)}}{\partial \tilde{\mathcal{C}}^2}\delta \tilde{\mathcal{C}}\hat{R}_{\mu\nu\rho\sigma}		
		\\& + \frac{\partial^2 \mathcal{L}_{(8)}}{\partial \mathcal{C}\partial \tilde{\mathcal{C}}}\left(\delta \mathcal{C}\hat{R}_{\mu\nu\rho\sigma} + \delta \tilde{\mathcal{C}} R_{\mu\nu\rho\sigma}\right)
		\,.
	\end{split}
\end{align}
Then, the right hand side of \eqref{eq:dispersion} in the large momentum limit simplifies to
\begin{align}
	\begin{split}
		&k^{\nu}k^{\sigma}\delta P^{(8)}_{\mu\nu\rho\sigma} \\&=
		8 \frac{\partial^2 \mathcal{L}_{(8)}}{\partial \mathcal{C}^2} (S_{\alpha\beta}h^{\alpha\beta}) S_{\mu\rho}
		+
		8 \frac{\partial^2 \mathcal{L}_{(8)}}{\partial \tilde{\mathcal{C}}^2} (\tilde{S}_{\alpha\beta}h^{\alpha\beta}) \tilde{S}_{\mu\rho}
		\\&
		+
		8 \frac{\partial^2 \mathcal{L}_{(8)}}{\partial \mathcal{C}\partial \tilde{\mathcal{C}}}\left((S_{\alpha\beta}h^{\alpha\beta})\tilde{S}_{\mu\rho} + (\tilde{S}_{\alpha\beta}h^{\alpha\beta})S_{\mu\rho}\right)\,,
	\end{split}
\end{align}
where we have used that $\delta \mathcal{C} = -4 S_{\mu\nu}h^{\mu\nu}$ and $\delta \tilde{\mathcal{C}} = -4 \tilde{S}_{\mu\nu}h^{\mu\nu}$, and that $k^{\nu}k^{\sigma}\delta R_{\mu\nu\rho\sigma}=k^{\nu}k^{\sigma}\delta \hat{R}_{\mu\nu\rho\sigma}=0$. This follows from the expression \eqref{eq:linearizedRiemann}, and from the facts that $h_{\mu\nu}$ is transverse and that $k_{\mu}$ is null at zeroth order in the higher-derivative corrections.

Finally, we contract \eqref{eq:dispersion} with $h^{\mu\nu}$ and impose the ansatz \eqref{eq:hansatz} to find the full expression for the dispersion relation. Once the dust settles, we have
\begin{align} \label{eq:keqquartic}
	\begin{split} 
		k^2&=
		32 \frac{\partial^2 \mathcal{L}_{(8)}}{\partial \mathcal{C}^2}(S^{\mu\nu}e_{\mu\nu})^2 + 32 \frac{\partial^2 \mathcal{L}_{(8)}}{\partial \tilde{\mathcal{C}}^2}(\tilde{S}^{\mu\nu}e_{\mu\nu})^2
		\\& + 64 \frac{\partial^2 \mathcal{L}_{(8)}}{\partial \mathcal{C}\partial\tilde{\mathcal{C}}}(S^{\mu\nu}e_{\mu\nu})(\tilde{S}^{\rho\sigma}e_{\rho\sigma})\,.
	\end{split}
\end{align}
We can explore theories where \eqref{eq:keqquartic} exhibits isospectral properties and is independent of $e_{\mu\nu}$. However, we find that the only theory has been given in \eqref{eq:Lspecial}.

\subsection{The dispersion relation for $\mathcal{L}_{(6)}=\mathcal{L}(\mathcal{I},\tilde{\mathcal{I}})$}
In a similar manner, we can analyze the corrections to the dispersion relation for a Lagrangian made of cubic order invariants $\mathcal{L}_{(6)}=\mathcal{L}(\mathcal{I},\tilde{\mathcal{I}})$. The variation of $\mathcal{L}_{(6)}$ with respect to the Riemann tensor takes the form
\begin{align}
	\begin{split}
	P^{(6)}_{\mu\nu\rho\sigma} &= 
	3\frac{\partial \mathcal{L}_{(6)}}{\partial\mathcal{I}} R_{\mu\nu\alpha\beta}R^{\alpha\beta}{}_{\rho\sigma} + \frac{\partial \mathcal{L}_{(6)}}{\partial \tilde{\mathcal{I}}}\mathcal{T}_{\mu\nu\rho\sigma}\,,
	\end{split}
\end{align}
with
\begin{align}
	\begin{split}
	\mathcal{T}_{\mu\nu\rho\sigma} &\equiv R_{\mu\nu}{}_{\alpha\beta}\tilde{R}^{\alpha\beta}{}_{\rho\sigma} +
	R_{\rho\sigma\alpha\beta} \tilde{R}_{\mu\nu}{}^{\alpha\beta}\\&+ R_{\mu\nu}{}_{\alpha\beta}\tilde{R}_{\rho\sigma}{}^{\alpha\beta}\,.
	\end{split}
\end{align}
At leading order in $k^{\mu}$, only \eqref{eq:leadingbehaviort3} needs to be taken into account whose variation explicitly is
\begin{align}
	\begin{split}
		\delta P^{(6)}_{\mu\nu\rho\sigma} &= 
		3\frac{\partial^2 \mathcal{L}_{(6)}}{\partial\mathcal{I}^2} \delta \mathcal{I} R_{\mu\nu}{}^{\alpha\beta}R_{\alpha\beta\rho\sigma} 
		\\& + 3\frac{\partial \mathcal{L}_{(6)}}{\partial \mathcal{I}}\left( \delta R_{\mu\nu}{}^{\alpha\beta}R_{\alpha\beta}{}_{\rho\sigma}+ R_{\mu\nu}{}^{\alpha\beta}\delta R_{\alpha\beta}{}_{\rho\sigma}\right)
		\\&
		+ \frac{\partial^2 \mathcal{L}_{(6)}}{\partial \mathcal{I}\partial \tilde{\mathcal{I}}}(3\delta \tilde{\mathcal{I}} R_{\mu\nu}{}^{\alpha\beta}R_{\alpha\beta\rho\sigma} + \delta \mathcal{I} \mathcal{T}_{\mu\nu\rho\sigma}) 
		\\& + 
		\frac{\partial^2 \mathcal{L}_{(6)}}{\partial \tilde{\mathcal{I}}^2}\delta \tilde{\mathcal{I}}\mathcal{T}_{\mu\nu\rho\sigma}
		+ \frac{\partial \mathcal{L}_{(6)}}{\partial \tilde{\mathcal{I}}}\delta \mathcal{T}_{\mu\nu\rho\sigma}\,.
	\end{split}
\end{align}
One can now proceed to produce the dispersion relation for the most general $\mathcal{L}_{(6)}$. However, to streamline our reasoning and show the non-isospectral behavior as found in the main text, we consider a Lagrangian linear in the cubic corrections, namely $\mathcal{L}_{(6)}=\lambda_{\rm ev} \mathcal{I} + \lambda_{\rm odd} \tilde{\mathcal{I}}$ where the variation of $\delta P^{(6)}_{\mu\nu\rho\sigma}$ takes a more compact form
\begin{align}
	\begin{split}
		\delta P^{(6)}_{\mu\nu\rho\sigma} &= 
		3 \lambda_{\text{ev}}\delta\left(  R_{\mu\nu}{}^{\alpha\beta}R_{\alpha\beta}{}_{\rho\sigma}\right) + \lambda_{\text{odd}}\delta \mathcal{T}_{\mu\nu\rho\sigma}\,.
	\end{split}
\end{align}
Once the dust settles, we find that at leading order $\mathcal{O}(k^4)$, the contribution to the dispersion relation vanishes, i.e.,
\begin{align}
	k^{\nu}k^{\sigma} \delta P^{(6)}_{\mu\nu\rho\sigma}  = \mathcal{O}(k^3)\,.
\end{align}
We can see this by using \eqref{eq:linearizedRiemann}. 
At the next order $\mathcal{O}(k^3)$, we can still neglect the first two terms in \eqref{eq:stressenergy}. This analysis can be done in Mathematica via the package xAct \cite{xAct}. We will further simplify the analysis by just considering the even parity term $\lambda_{\text{ev}}\mathcal{I}$. We find the dispersion relation takes the compact form
\begin{align} \label{eq:kleading}
	k^2 h^{\mu\nu}h_{\mu\nu} = 12 \lambda_{\rm ev} i  h_{\delta}{}^{\gamma}h^{\delta \epsilon}k^{\alpha}k^{\beta}k^{\rho}\nabla_{\rho}R_{\alpha\epsilon\beta\gamma} + \mathcal{O}(k^2)\,,
\end{align}
Once again, \eqref{eq:kleading} is zero at this order, which can be shown using the identity
\begin{align}
	0 &= e^{[\alpha}{}_{\gamma} e^{\gamma}{}_{\beta} k^{\tau|} k_{\delta} k^{\kappa} \nabla^{|\delta} R_{\alpha \tau \kappa}{}{}^{\beta]}\,.
\end{align}
The dispersion relation for this theory is therefore of $\mathcal{O}(k^2)$. At this order, we must also include \eqref{eq:leadingbehaviort1} and \eqref{eq:leadingbehaviort2}. What we find are terms of different structure, such as ones that include the derivatives of $e_{\mu\nu}$. Unlike the quartic theory \eqref{eq:Lspecial}, even after simplifying and using known identities,
the dispersion relation does depend on the polarization in a very complicated way. This is in agreement with the polarization dependence found in the potential in the eikonal limit \eqref{cubicdV} which in fact scales as $l^2$.

\subsection{Regime of validity}
We would like to make a few remarks on the regime of validity of both the EFT and the geometric optics limit. Let us denote $\ell$ as the length scale of new physics (in terms the coupling constants we have $\lambda_{\rm ev,odd}\sim \ell^4$ and $\epsilon_{i}\sim \ell^6$) and $1/L^2$ as the scale of the background curvature of the spacetime. The validity of the EFT requires that $\ell/L \ll 1$, since higher-curvature corrections scale with powers of $\ell/L$.  On the other hand, the geometric optics limit requires that the wavelength $\lambda$ is small relative to the size of the curvature, \textit{i.e}., $\lambda\ll  L$. 
However, the wavelength cannot be too small in order to remain within the regime of validity of the EFT. It has been previously noticed that if one assumes the natural cutoff  $\lambda\gg \ell$, then the higher-derivative corrections to the dispersion relation are always smaller than subleading terms in the geometric optics expansion of GR \cite{Goon:2016une,deRham:2020zyh,Reall:2021voz}. This means that the higher-derivative effects on the light cone of GWs would not be ``resolvable'' within EFT, since they would be smaller than finite size effects. Nevertheless, the dispersion relation is at the end of the day an statement on the form of the equations of motion of the theory, and this has physical implications \cite{Reall:2021voz}. For instance, as we have seen, non-birefringence leads to isospectrality of the higher-derivative corrections to the eikonal QNMs, which is an observable property independently of higher-derivative corrections being smaller or larger then finite size effects in the eikonal expansion. 

Additionally, we would like to point out that the bound $\lambda\gg \ell$ may be too strong for the theories that we consider. In fact, a guiding principle to determine the validity of the EFT should be that the corrections to GR remain small. For instance, in the case of the quartic dispersion relation \req{eq:keqquartic}, the corrections to GR remain small as long as $\lambda\gg \ell^3/L^2$. If one takes this as the smallest wavelength resolvable by the EFT, then there is an interval in which the higher-derivative corrections are larger than finite size effects: $\ell^3/L^2\ll \lambda\ll \ell^2/L$. In this regime, higher-derivative corrections to the dispersion relation would be resolvable.

\bibliography{Gravities}

\begin{thebibliography}{62}%
\makeatletter
\providecommand \@ifxundefined [1]{%
 \@ifx{#1\undefined}
}%
\providecommand \@ifnum [1]{%
 \ifnum #1\expandafter \@firstoftwo
 \else \expandafter \@secondoftwo
 \fi
}%
\providecommand \@ifx [1]{%
 \ifx #1\expandafter \@firstoftwo
 \else \expandafter \@secondoftwo
 \fi
}%
\providecommand \natexlab [1]{#1}%
\providecommand \enquote  [1]{``#1''}%
\providecommand \bibnamefont  [1]{#1}%
\providecommand \bibfnamefont [1]{#1}%
\providecommand \citenamefont [1]{#1}%
\providecommand \href@noop [0]{\@secondoftwo}%
\providecommand \href [0]{\begingroup \@sanitize@url \@href}%
\providecommand \@href[1]{\@@startlink{#1}\@@href}%
\providecommand \@@href[1]{\endgroup#1\@@endlink}%
\providecommand \@sanitize@url [0]{\catcode `\\12\catcode `\$12\catcode
  `\&12\catcode `\#12\catcode `\^12\catcode `\_12\catcode `\%12\relax}%
\providecommand \@@startlink[1]{}%
\providecommand \@@endlink[0]{}%
\providecommand \url  [0]{\begingroup\@sanitize@url \@url }%
\providecommand \@url [1]{\endgroup\@href {#1}{\urlprefix }}%
\providecommand \urlprefix  [0]{URL }%
\providecommand \Eprint [0]{\href }%
\providecommand \doibase [0]{https://doi.org/}%
\providecommand \selectlanguage [0]{\@gobble}%
\providecommand \bibinfo  [0]{\@secondoftwo}%
\providecommand \bibfield  [0]{\@secondoftwo}%
\providecommand \translation [1]{[#1]}%
\providecommand \BibitemOpen [0]{}%
\providecommand \bibitemStop [0]{}%
\providecommand \bibitemNoStop [0]{.\EOS\space}%
\providecommand \EOS [0]{\spacefactor3000\relax}%
\providecommand \BibitemShut  [1]{\csname bibitem#1\endcsname}%
\let\auto@bib@innerbib\@empty
\bibitem [{\citenamefont {Gross}\ and\ \citenamefont
  {Witten}(1986)}]{Gross:1986iv}%
  \BibitemOpen
  \bibfield  {author} {\bibinfo {author} {\bibfnamefont {D.~J.}\ \bibnamefont
  {Gross}}\ and\ \bibinfo {author} {\bibfnamefont {E.}~\bibnamefont {Witten}},\
  }\bibfield  {title} {\bibinfo {title} {{Superstring Modifications of
  Einstein's Equations}},\ }\href
  {https://doi.org/10.1016/0550-3213(86)90429-3} {\bibfield  {journal}
  {\bibinfo  {journal} {Nucl. Phys. B}\ }\textbf {\bibinfo {volume} {277}},\
  \bibinfo {pages} {1} (\bibinfo {year} {1986})}\BibitemShut {NoStop}%
\bibitem [{\citenamefont {Gross}\ and\ \citenamefont
  {Sloan}(1987)}]{Gross:1986mw}%
  \BibitemOpen
  \bibfield  {author} {\bibinfo {author} {\bibfnamefont {D.~J.}\ \bibnamefont
  {Gross}}\ and\ \bibinfo {author} {\bibfnamefont {J.~H.}\ \bibnamefont
  {Sloan}},\ }\bibfield  {title} {\bibinfo {title} {{The Quartic Effective
  Action for the Heterotic String}},\ }\href
  {https://doi.org/10.1016/0550-3213(87)90465-2} {\bibfield  {journal}
  {\bibinfo  {journal} {Nucl. Phys. B}\ }\textbf {\bibinfo {volume} {291}},\
  \bibinfo {pages} {41} (\bibinfo {year} {1987})}\BibitemShut {NoStop}%
\bibitem [{\citenamefont {Bergshoeff}\ and\ \citenamefont
  {de~Roo}(1989)}]{Bergshoeff:1989de}%
  \BibitemOpen
  \bibfield  {author} {\bibinfo {author} {\bibfnamefont {E.~A.}\ \bibnamefont
  {Bergshoeff}}\ and\ \bibinfo {author} {\bibfnamefont {M.}~\bibnamefont
  {de~Roo}},\ }\bibfield  {title} {\bibinfo {title} {{The Quartic Effective
  Action of the Heterotic String and Supersymmetry}},\ }\href
  {https://doi.org/10.1016/0550-3213(89)90336-2} {\bibfield  {journal}
  {\bibinfo  {journal} {Nucl. Phys. B}\ }\textbf {\bibinfo {volume} {328}},\
  \bibinfo {pages} {439} (\bibinfo {year} {1989})}\BibitemShut {NoStop}%
\bibitem [{\citenamefont {Endlich}\ \emph {et~al.}(2017)\citenamefont
  {Endlich}, \citenamefont {Gorbenko}, \citenamefont {Huang},\ and\
  \citenamefont {Senatore}}]{Endlich:2017tqa}%
  \BibitemOpen
  \bibfield  {author} {\bibinfo {author} {\bibfnamefont {S.}~\bibnamefont
  {Endlich}}, \bibinfo {author} {\bibfnamefont {V.}~\bibnamefont {Gorbenko}},
  \bibinfo {author} {\bibfnamefont {J.}~\bibnamefont {Huang}},\ and\ \bibinfo
  {author} {\bibfnamefont {L.}~\bibnamefont {Senatore}},\ }\bibfield  {title}
  {\bibinfo {title} {{An effective formalism for testing extensions to General
  Relativity with gravitational waves}},\ }\href
  {https://doi.org/10.1007/JHEP09(2017)122} {\bibfield  {journal} {\bibinfo
  {journal} {JHEP}\ }\textbf {\bibinfo {volume} {09}},\ \bibinfo {pages}
  {122}},\ \Eprint {https://arxiv.org/abs/1704.01590} {arXiv:1704.01590
  [gr-qc]} \BibitemShut {NoStop}%
\bibitem [{\citenamefont {Cano}\ and\ \citenamefont
  {Ruip\'erez}(2019)}]{Cano:2019ore}%
  \BibitemOpen
  \bibfield  {author} {\bibinfo {author} {\bibfnamefont {P.~A.}\ \bibnamefont
  {Cano}}\ and\ \bibinfo {author} {\bibfnamefont {A.}~\bibnamefont
  {Ruip\'erez}},\ }\bibfield  {title} {\bibinfo {title} {{Leading
  higher-derivative corrections to Kerr geometry}},\ }\href
  {https://doi.org/10.1007/JHEP05(2019)189} {\bibfield  {journal} {\bibinfo
  {journal} {JHEP}\ }\textbf {\bibinfo {volume} {05}},\ \bibinfo {pages}
  {189}},\ \bibinfo {note} {[Erratum: JHEP 03, 187 (2020)]},\ \Eprint
  {https://arxiv.org/abs/1901.01315} {arXiv:1901.01315 [gr-qc]} \BibitemShut
  {NoStop}%
\bibitem [{\citenamefont {Ruhdorfer}\ \emph {et~al.}(2020)\citenamefont
  {Ruhdorfer}, \citenamefont {Serra},\ and\ \citenamefont
  {Weiler}}]{Ruhdorfer:2019qmk}%
  \BibitemOpen
  \bibfield  {author} {\bibinfo {author} {\bibfnamefont {M.}~\bibnamefont
  {Ruhdorfer}}, \bibinfo {author} {\bibfnamefont {J.}~\bibnamefont {Serra}},\
  and\ \bibinfo {author} {\bibfnamefont {A.}~\bibnamefont {Weiler}},\
  }\bibfield  {title} {\bibinfo {title} {{Effective Field Theory of Gravity to
  All Orders}},\ }\href {https://doi.org/10.1007/JHEP05(2020)083} {\bibfield
  {journal} {\bibinfo  {journal} {JHEP}\ }\textbf {\bibinfo {volume} {05}},\
  \bibinfo {pages} {083}},\ \Eprint {https://arxiv.org/abs/1908.08050}
  {arXiv:1908.08050 [hep-ph]} \BibitemShut {NoStop}%
\bibitem [{\citenamefont {Adams}\ \emph {et~al.}(2006)\citenamefont {Adams},
  \citenamefont {Arkani-Hamed}, \citenamefont {Dubovsky}, \citenamefont
  {Nicolis},\ and\ \citenamefont {Rattazzi}}]{Adams:2006sv}%
  \BibitemOpen
  \bibfield  {author} {\bibinfo {author} {\bibfnamefont {A.}~\bibnamefont
  {Adams}}, \bibinfo {author} {\bibfnamefont {N.}~\bibnamefont {Arkani-Hamed}},
  \bibinfo {author} {\bibfnamefont {S.}~\bibnamefont {Dubovsky}}, \bibinfo
  {author} {\bibfnamefont {A.}~\bibnamefont {Nicolis}},\ and\ \bibinfo {author}
  {\bibfnamefont {R.}~\bibnamefont {Rattazzi}},\ }\bibfield  {title} {\bibinfo
  {title} {{Causality, analyticity and an IR obstruction to UV completion}},\
  }\href {https://doi.org/10.1088/1126-6708/2006/10/014} {\bibfield  {journal}
  {\bibinfo  {journal} {JHEP}\ }\textbf {\bibinfo {volume} {10}},\ \bibinfo
  {pages} {014}},\ \Eprint {https://arxiv.org/abs/hep-th/0602178}
  {arXiv:hep-th/0602178} \BibitemShut {NoStop}%
\bibitem [{\citenamefont {Gruzinov}\ and\ \citenamefont
  {Kleban}(2007)}]{Gruzinov:2006ie}%
  \BibitemOpen
  \bibfield  {author} {\bibinfo {author} {\bibfnamefont {A.}~\bibnamefont
  {Gruzinov}}\ and\ \bibinfo {author} {\bibfnamefont {M.}~\bibnamefont
  {Kleban}},\ }\bibfield  {title} {\bibinfo {title} {{Causality Constrains
  Higher Curvature Corrections to Gravity}},\ }\href
  {https://doi.org/10.1088/0264-9381/24/13/N02} {\bibfield  {journal} {\bibinfo
   {journal} {Class. Quant. Grav.}\ }\textbf {\bibinfo {volume} {24}},\
  \bibinfo {pages} {3521} (\bibinfo {year} {2007})},\ \Eprint
  {https://arxiv.org/abs/hep-th/0612015} {arXiv:hep-th/0612015} \BibitemShut
  {NoStop}%
\bibitem [{\citenamefont {Brigante}\ \emph {et~al.}(2008)\citenamefont
  {Brigante}, \citenamefont {Liu}, \citenamefont {Myers}, \citenamefont
  {Shenker},\ and\ \citenamefont {Yaida}}]{Brigante:2008gz}%
  \BibitemOpen
  \bibfield  {author} {\bibinfo {author} {\bibfnamefont {M.}~\bibnamefont
  {Brigante}}, \bibinfo {author} {\bibfnamefont {H.}~\bibnamefont {Liu}},
  \bibinfo {author} {\bibfnamefont {R.~C.}\ \bibnamefont {Myers}}, \bibinfo
  {author} {\bibfnamefont {S.}~\bibnamefont {Shenker}},\ and\ \bibinfo {author}
  {\bibfnamefont {S.}~\bibnamefont {Yaida}},\ }\bibfield  {title} {\bibinfo
  {title} {{The Viscosity Bound and Causality Violation}},\ }\href
  {https://doi.org/10.1103/PhysRevLett.100.191601} {\bibfield  {journal}
  {\bibinfo  {journal} {Phys. Rev. Lett.}\ }\textbf {\bibinfo {volume} {100}},\
  \bibinfo {pages} {191601} (\bibinfo {year} {2008})},\ \Eprint
  {https://arxiv.org/abs/0802.3318} {arXiv:0802.3318 [hep-th]} \BibitemShut
  {NoStop}%
\bibitem [{\citenamefont {Camanho}\ \emph {et~al.}(2016)\citenamefont
  {Camanho}, \citenamefont {Edelstein}, \citenamefont {Maldacena},\ and\
  \citenamefont {Zhiboedov}}]{Camanho:2014apa}%
  \BibitemOpen
  \bibfield  {author} {\bibinfo {author} {\bibfnamefont {X.~O.}\ \bibnamefont
  {Camanho}}, \bibinfo {author} {\bibfnamefont {J.~D.}\ \bibnamefont
  {Edelstein}}, \bibinfo {author} {\bibfnamefont {J.}~\bibnamefont
  {Maldacena}},\ and\ \bibinfo {author} {\bibfnamefont {A.}~\bibnamefont
  {Zhiboedov}},\ }\bibfield  {title} {\bibinfo {title} {{Causality Constraints
  on Corrections to the Graviton Three-Point Coupling}},\ }\href
  {https://doi.org/10.1007/JHEP02(2016)020} {\bibfield  {journal} {\bibinfo
  {journal} {JHEP}\ }\textbf {\bibinfo {volume} {02}},\ \bibinfo {pages}
  {020}},\ \Eprint {https://arxiv.org/abs/1407.5597} {arXiv:1407.5597 [hep-th]}
  \BibitemShut {NoStop}%
\bibitem [{\citenamefont {Goon}\ and\ \citenamefont
  {Hinterbichler}(2017)}]{Goon:2016une}%
  \BibitemOpen
  \bibfield  {author} {\bibinfo {author} {\bibfnamefont {G.}~\bibnamefont
  {Goon}}\ and\ \bibinfo {author} {\bibfnamefont {K.}~\bibnamefont
  {Hinterbichler}},\ }\bibfield  {title} {\bibinfo {title} {{Superluminality,
  black holes and EFT}},\ }\href {https://doi.org/10.1007/JHEP02(2017)134}
  {\bibfield  {journal} {\bibinfo  {journal} {JHEP}\ }\textbf {\bibinfo
  {volume} {02}},\ \bibinfo {pages} {134}},\ \Eprint
  {https://arxiv.org/abs/1609.00723} {arXiv:1609.00723 [hep-th]} \BibitemShut
  {NoStop}%
\bibitem [{\citenamefont {de~Rham}\ \emph {et~al.}(2020)\citenamefont
  {de~Rham}, \citenamefont {Francfort},\ and\ \citenamefont
  {Zhang}}]{deRham:2020ejn}%
  \BibitemOpen
  \bibfield  {author} {\bibinfo {author} {\bibfnamefont {C.}~\bibnamefont
  {de~Rham}}, \bibinfo {author} {\bibfnamefont {J.}~\bibnamefont {Francfort}},\
  and\ \bibinfo {author} {\bibfnamefont {J.}~\bibnamefont {Zhang}},\ }\bibfield
   {title} {\bibinfo {title} {{Black Hole Gravitational Waves in the Effective
  Field Theory of Gravity}},\ }\href
  {https://doi.org/10.1103/PhysRevD.102.024079} {\bibfield  {journal} {\bibinfo
   {journal} {Phys. Rev. D}\ }\textbf {\bibinfo {volume} {102}},\ \bibinfo
  {pages} {024079} (\bibinfo {year} {2020})},\ \Eprint
  {https://arxiv.org/abs/2005.13923} {arXiv:2005.13923 [hep-th]} \BibitemShut
  {NoStop}%
\bibitem [{\citenamefont {de~Rham}\ and\ \citenamefont
  {Tolley}(2020)}]{deRham:2020zyh}%
  \BibitemOpen
  \bibfield  {author} {\bibinfo {author} {\bibfnamefont {C.}~\bibnamefont
  {de~Rham}}\ and\ \bibinfo {author} {\bibfnamefont {A.~J.}\ \bibnamefont
  {Tolley}},\ }\bibfield  {title} {\bibinfo {title} {{Causality in curved
  spacetimes: The speed of light and gravity}},\ }\href
  {https://doi.org/10.1103/PhysRevD.102.084048} {\bibfield  {journal} {\bibinfo
   {journal} {Phys. Rev. D}\ }\textbf {\bibinfo {volume} {102}},\ \bibinfo
  {pages} {084048} (\bibinfo {year} {2020})},\ \Eprint
  {https://arxiv.org/abs/2007.01847} {arXiv:2007.01847 [hep-th]} \BibitemShut
  {NoStop}%
\bibitem [{\citenamefont {Accettulli~Huber}\ \emph {et~al.}(2021)\citenamefont
  {Accettulli~Huber}, \citenamefont {Brandhuber}, \citenamefont {De~Angelis},\
  and\ \citenamefont {Travaglini}}]{AccettulliHuber:2020dal}%
  \BibitemOpen
  \bibfield  {author} {\bibinfo {author} {\bibfnamefont {M.}~\bibnamefont
  {Accettulli~Huber}}, \bibinfo {author} {\bibfnamefont {A.}~\bibnamefont
  {Brandhuber}}, \bibinfo {author} {\bibfnamefont {S.}~\bibnamefont
  {De~Angelis}},\ and\ \bibinfo {author} {\bibfnamefont {G.}~\bibnamefont
  {Travaglini}},\ }\bibfield  {title} {\bibinfo {title} {{From amplitudes to
  gravitational radiation with cubic interactions and tidal effects}},\ }\href
  {https://doi.org/10.1103/PhysRevD.103.045015} {\bibfield  {journal} {\bibinfo
   {journal} {Phys. Rev. D}\ }\textbf {\bibinfo {volume} {103}},\ \bibinfo
  {pages} {045015} (\bibinfo {year} {2021})},\ \Eprint
  {https://arxiv.org/abs/2012.06548} {arXiv:2012.06548 [hep-th]} \BibitemShut
  {NoStop}%
\bibitem [{\citenamefont {de~Rham}\ \emph {et~al.}(2022)\citenamefont
  {de~Rham}, \citenamefont {Tolley},\ and\ \citenamefont
  {Zhang}}]{deRham:2021bll}%
  \BibitemOpen
  \bibfield  {author} {\bibinfo {author} {\bibfnamefont {C.}~\bibnamefont
  {de~Rham}}, \bibinfo {author} {\bibfnamefont {A.~J.}\ \bibnamefont
  {Tolley}},\ and\ \bibinfo {author} {\bibfnamefont {J.}~\bibnamefont
  {Zhang}},\ }\bibfield  {title} {\bibinfo {title} {{Causality Constraints on
  Gravitational Effective Field Theories}},\ }\href
  {https://doi.org/10.1103/PhysRevLett.128.131102} {\bibfield  {journal}
  {\bibinfo  {journal} {Phys. Rev. Lett.}\ }\textbf {\bibinfo {volume} {128}},\
  \bibinfo {pages} {131102} (\bibinfo {year} {2022})},\ \Eprint
  {https://arxiv.org/abs/2112.05054} {arXiv:2112.05054 [gr-qc]} \BibitemShut
  {NoStop}%
\bibitem [{\citenamefont {Edelstein}\ \emph {et~al.}(2021)\citenamefont
  {Edelstein}, \citenamefont {Ghosh}, \citenamefont {Laddha},\ and\
  \citenamefont {Sarkar}}]{Edelstein:2021jyu}%
  \BibitemOpen
  \bibfield  {author} {\bibinfo {author} {\bibfnamefont {J.~D.}\ \bibnamefont
  {Edelstein}}, \bibinfo {author} {\bibfnamefont {R.}~\bibnamefont {Ghosh}},
  \bibinfo {author} {\bibfnamefont {A.}~\bibnamefont {Laddha}},\ and\ \bibinfo
  {author} {\bibfnamefont {S.}~\bibnamefont {Sarkar}},\ }\bibfield  {title}
  {\bibinfo {title} {{Causality constraints in Quadratic Gravity}},\ }\href
  {https://doi.org/10.1007/JHEP09(2021)150} {\bibfield  {journal} {\bibinfo
  {journal} {JHEP}\ }\textbf {\bibinfo {volume} {09}},\ \bibinfo {pages}
  {150}},\ \Eprint {https://arxiv.org/abs/2107.07424} {arXiv:2107.07424
  [hep-th]} \BibitemShut {NoStop}%
\bibitem [{\citenamefont {Caron-Huot}\ \emph {et~al.}(2023)\citenamefont
  {Caron-Huot}, \citenamefont {Li}, \citenamefont {Parra-Martinez},\ and\
  \citenamefont {Simmons-Duffin}}]{Caron-Huot:2022ugt}%
  \BibitemOpen
  \bibfield  {author} {\bibinfo {author} {\bibfnamefont {S.}~\bibnamefont
  {Caron-Huot}}, \bibinfo {author} {\bibfnamefont {Y.-Z.}\ \bibnamefont {Li}},
  \bibinfo {author} {\bibfnamefont {J.}~\bibnamefont {Parra-Martinez}},\ and\
  \bibinfo {author} {\bibfnamefont {D.}~\bibnamefont {Simmons-Duffin}},\
  }\bibfield  {title} {\bibinfo {title} {{Causality constraints on corrections
  to Einstein gravity}},\ }\href {https://doi.org/10.1007/JHEP05(2023)122}
  {\bibfield  {journal} {\bibinfo  {journal} {JHEP}\ }\textbf {\bibinfo
  {volume} {05}},\ \bibinfo {pages} {122}},\ \Eprint
  {https://arxiv.org/abs/2201.06602} {arXiv:2201.06602 [hep-th]} \BibitemShut
  {NoStop}%
\bibitem [{\citenamefont {Cremonini}\ \emph {et~al.}(2024)\citenamefont
  {Cremonini}, \citenamefont {McPeak},\ and\ \citenamefont
  {Tang}}]{Cremonini:2023epg}%
  \BibitemOpen
  \bibfield  {author} {\bibinfo {author} {\bibfnamefont {S.}~\bibnamefont
  {Cremonini}}, \bibinfo {author} {\bibfnamefont {B.}~\bibnamefont {McPeak}},\
  and\ \bibinfo {author} {\bibfnamefont {Y.}~\bibnamefont {Tang}},\ }\bibfield
  {title} {\bibinfo {title} {{Electric shocks: bounding Einstein-Maxwell theory
  with time delays on boosted RN backgrounds}},\ }\href
  {https://doi.org/10.1007/JHEP05(2024)192} {\bibfield  {journal} {\bibinfo
  {journal} {JHEP}\ }\textbf {\bibinfo {volume} {05}},\ \bibinfo {pages}
  {192}},\ \Eprint {https://arxiv.org/abs/2312.17328} {arXiv:2312.17328
  [hep-th]} \BibitemShut {NoStop}%
\bibitem [{\citenamefont {Vafa}(2005)}]{Vafa:2005ui}%
  \BibitemOpen
  \bibfield  {author} {\bibinfo {author} {\bibfnamefont {C.}~\bibnamefont
  {Vafa}},\ }\bibfield  {title} {\bibinfo {title} {{The String landscape and
  the swampland}},\ }\href@noop {} {\  (\bibinfo {year} {2005})},\ \Eprint
  {https://arxiv.org/abs/hep-th/0509212} {arXiv:hep-th/0509212} \BibitemShut
  {NoStop}%
\bibitem [{\citenamefont {Berti}\ \emph {et~al.}(2009)\citenamefont {Berti},
  \citenamefont {Cardoso},\ and\ \citenamefont {Starinets}}]{Berti:2009kk}%
  \BibitemOpen
  \bibfield  {author} {\bibinfo {author} {\bibfnamefont {E.}~\bibnamefont
  {Berti}}, \bibinfo {author} {\bibfnamefont {V.}~\bibnamefont {Cardoso}},\
  and\ \bibinfo {author} {\bibfnamefont {A.~O.}\ \bibnamefont {Starinets}},\
  }\bibfield  {title} {\bibinfo {title} {{Quasinormal modes of black holes and
  black branes}},\ }\href {https://doi.org/10.1088/0264-9381/26/16/163001}
  {\bibfield  {journal} {\bibinfo  {journal} {Class. Quant. Grav.}\ }\textbf
  {\bibinfo {volume} {26}},\ \bibinfo {pages} {163001} (\bibinfo {year}
  {2009})},\ \Eprint {https://arxiv.org/abs/0905.2975} {arXiv:0905.2975
  [gr-qc]} \BibitemShut {NoStop}%
\bibitem [{\citenamefont {Cardoso}\ and\ \citenamefont
  {Gualtieri}(2009)}]{Cardoso:2009pk}%
  \BibitemOpen
  \bibfield  {author} {\bibinfo {author} {\bibfnamefont {V.}~\bibnamefont
  {Cardoso}}\ and\ \bibinfo {author} {\bibfnamefont {L.}~\bibnamefont
  {Gualtieri}},\ }\bibfield  {title} {\bibinfo {title} {{Perturbations of
  Schwarzschild black holes in Dynamical Chern-Simons modified gravity}},\
  }\href {https://doi.org/10.1103/PhysRevD.81.089903,
  10.1103/PhysRevD.80.064008} {\bibfield  {journal} {\bibinfo  {journal} {Phys.
  Rev.}\ }\textbf {\bibinfo {volume} {D80}},\ \bibinfo {pages} {064008}
  (\bibinfo {year} {2009})},\ \bibinfo {note} {[Erratum: Phys.
  Rev.D81,089903(2010)]},\ \Eprint {https://arxiv.org/abs/0907.5008}
  {arXiv:0907.5008 [gr-qc]} \BibitemShut {NoStop}%
\bibitem [{\citenamefont {Blázquez-Salcedo}\ \emph {et~al.}(2016)\citenamefont
  {Blázquez-Salcedo}, \citenamefont {Macedo}, \citenamefont {Cardoso},
  \citenamefont {Ferrari}, \citenamefont {Gualtieri}, \citenamefont {Khoo},
  \citenamefont {Kunz},\ and\ \citenamefont {Pani}}]{Blazquez-Salcedo:2016enn}%
  \BibitemOpen
  \bibfield  {author} {\bibinfo {author} {\bibfnamefont {J.~L.}\ \bibnamefont
  {Blázquez-Salcedo}}, \bibinfo {author} {\bibfnamefont {C.~F.~B.}\
  \bibnamefont {Macedo}}, \bibinfo {author} {\bibfnamefont {V.}~\bibnamefont
  {Cardoso}}, \bibinfo {author} {\bibfnamefont {V.}~\bibnamefont {Ferrari}},
  \bibinfo {author} {\bibfnamefont {L.}~\bibnamefont {Gualtieri}}, \bibinfo
  {author} {\bibfnamefont {F.~S.}\ \bibnamefont {Khoo}}, \bibinfo {author}
  {\bibfnamefont {J.}~\bibnamefont {Kunz}},\ and\ \bibinfo {author}
  {\bibfnamefont {P.}~\bibnamefont {Pani}},\ }\bibfield  {title} {\bibinfo
  {title} {{Perturbed black holes in Einstein-dilaton-Gauss-Bonnet gravity:
  Stability, ringdown, and gravitational-wave emission}},\ }\href
  {https://doi.org/10.1103/PhysRevD.94.104024} {\bibfield  {journal} {\bibinfo
  {journal} {Phys. Rev.}\ }\textbf {\bibinfo {volume} {D94}},\ \bibinfo {pages}
  {104024} (\bibinfo {year} {2016})},\ \Eprint
  {https://arxiv.org/abs/1609.01286} {arXiv:1609.01286 [gr-qc]} \BibitemShut
  {NoStop}%
\bibitem [{\citenamefont {Cardoso}\ \emph {et~al.}(2018)\citenamefont
  {Cardoso}, \citenamefont {Kimura}, \citenamefont {Maselli},\ and\
  \citenamefont {Senatore}}]{Cardoso:2018ptl}%
  \BibitemOpen
  \bibfield  {author} {\bibinfo {author} {\bibfnamefont {V.}~\bibnamefont
  {Cardoso}}, \bibinfo {author} {\bibfnamefont {M.}~\bibnamefont {Kimura}},
  \bibinfo {author} {\bibfnamefont {A.}~\bibnamefont {Maselli}},\ and\ \bibinfo
  {author} {\bibfnamefont {L.}~\bibnamefont {Senatore}},\ }\bibfield  {title}
  {\bibinfo {title} {{Black Holes in an Effective Field Theory Extension of
  General Relativity}},\ }\href
  {https://doi.org/10.1103/PhysRevLett.121.251105} {\bibfield  {journal}
  {\bibinfo  {journal} {Phys. Rev. Lett.}\ }\textbf {\bibinfo {volume} {121}},\
  \bibinfo {pages} {251105} (\bibinfo {year} {2018})},\ \Eprint
  {https://arxiv.org/abs/1808.08962} {arXiv:1808.08962 [gr-qc]} \BibitemShut
  {NoStop}%
\bibitem [{\citenamefont {Bhattacharyya}\ and\ \citenamefont
  {Shankaranarayanan}(2019)}]{Bhattacharyya:2018hsj}%
  \BibitemOpen
  \bibfield  {author} {\bibinfo {author} {\bibfnamefont {S.}~\bibnamefont
  {Bhattacharyya}}\ and\ \bibinfo {author} {\bibfnamefont {S.}~\bibnamefont
  {Shankaranarayanan}},\ }\bibfield  {title} {\bibinfo {title} {{Distinguishing
  general relativity from Chern-Simons gravity using gravitational wave
  polarizations}},\ }\href {https://doi.org/10.1103/PhysRevD.100.024022}
  {\bibfield  {journal} {\bibinfo  {journal} {Phys. Rev. D}\ }\textbf {\bibinfo
  {volume} {100}},\ \bibinfo {pages} {024022} (\bibinfo {year} {2019})},\
  \Eprint {https://arxiv.org/abs/1812.00187} {arXiv:1812.00187 [gr-qc]}
  \BibitemShut {NoStop}%
\bibitem [{\citenamefont {Moura}\ and\ \citenamefont
  {Rodrigues}(2021)}]{Moura:2021nuh}%
  \BibitemOpen
  \bibfield  {author} {\bibinfo {author} {\bibfnamefont {F.}~\bibnamefont
  {Moura}}\ and\ \bibinfo {author} {\bibfnamefont {J.~a.}\ \bibnamefont
  {Rodrigues}},\ }\bibfield  {title} {\bibinfo {title} {{Asymptotic quasinormal
  modes of string-theoretical d-dimensional black holes}},\ }\href
  {https://doi.org/10.1007/JHEP08(2021)078} {\bibfield  {journal} {\bibinfo
  {journal} {JHEP}\ }\textbf {\bibinfo {volume} {08}},\ \bibinfo {pages}
  {078}},\ \Eprint {https://arxiv.org/abs/2105.02616} {arXiv:2105.02616
  [hep-th]} \BibitemShut {NoStop}%
\bibitem [{\citenamefont {Cano}\ \emph
  {et~al.}(2022{\natexlab{a}})\citenamefont {Cano}, \citenamefont {Fransen},
  \citenamefont {Hertog},\ and\ \citenamefont {Maenaut}}]{Cano:2021myl}%
  \BibitemOpen
  \bibfield  {author} {\bibinfo {author} {\bibfnamefont {P.~A.}\ \bibnamefont
  {Cano}}, \bibinfo {author} {\bibfnamefont {K.}~\bibnamefont {Fransen}},
  \bibinfo {author} {\bibfnamefont {T.}~\bibnamefont {Hertog}},\ and\ \bibinfo
  {author} {\bibfnamefont {S.}~\bibnamefont {Maenaut}},\ }\bibfield  {title}
  {\bibinfo {title} {{Gravitational ringing of rotating black holes in
  higher-derivative gravity}},\ }\href
  {https://doi.org/10.1103/PhysRevD.105.024064} {\bibfield  {journal} {\bibinfo
   {journal} {Phys. Rev. D}\ }\textbf {\bibinfo {volume} {105}},\ \bibinfo
  {pages} {024064} (\bibinfo {year} {2022}{\natexlab{a}})},\ \Eprint
  {https://arxiv.org/abs/2110.11378} {arXiv:2110.11378 [gr-qc]} \BibitemShut
  {NoStop}%
\bibitem [{\citenamefont {Cano}\ \emph
  {et~al.}(2023{\natexlab{a}})\citenamefont {Cano}, \citenamefont {Fransen},
  \citenamefont {Hertog},\ and\ \citenamefont {Maenaut}}]{Cano:2023tmv}%
  \BibitemOpen
  \bibfield  {author} {\bibinfo {author} {\bibfnamefont {P.~A.}\ \bibnamefont
  {Cano}}, \bibinfo {author} {\bibfnamefont {K.}~\bibnamefont {Fransen}},
  \bibinfo {author} {\bibfnamefont {T.}~\bibnamefont {Hertog}},\ and\ \bibinfo
  {author} {\bibfnamefont {S.}~\bibnamefont {Maenaut}},\ }\bibfield  {title}
  {\bibinfo {title} {{Universal Teukolsky equations and black hole
  perturbations in higher-derivative gravity}},\ }\href
  {https://doi.org/10.1103/PhysRevD.108.024040} {\bibfield  {journal} {\bibinfo
   {journal} {Phys. Rev. D}\ }\textbf {\bibinfo {volume} {108}},\ \bibinfo
  {pages} {024040} (\bibinfo {year} {2023}{\natexlab{a}})},\ \Eprint
  {https://arxiv.org/abs/2304.02663} {arXiv:2304.02663 [gr-qc]} \BibitemShut
  {NoStop}%
\bibitem [{\citenamefont {Li}\ \emph {et~al.}(2024)\citenamefont {Li},
  \citenamefont {Hussain}, \citenamefont {Wagle}, \citenamefont {Chen},
  \citenamefont {Yunes},\ and\ \citenamefont {Zimmerman}}]{Li:2023ulk}%
  \BibitemOpen
  \bibfield  {author} {\bibinfo {author} {\bibfnamefont {D.}~\bibnamefont
  {Li}}, \bibinfo {author} {\bibfnamefont {A.}~\bibnamefont {Hussain}},
  \bibinfo {author} {\bibfnamefont {P.}~\bibnamefont {Wagle}}, \bibinfo
  {author} {\bibfnamefont {Y.}~\bibnamefont {Chen}}, \bibinfo {author}
  {\bibfnamefont {N.}~\bibnamefont {Yunes}},\ and\ \bibinfo {author}
  {\bibfnamefont {A.}~\bibnamefont {Zimmerman}},\ }\bibfield  {title} {\bibinfo
  {title} {{Isospectrality breaking in the Teukolsky formalism}},\ }\href
  {https://doi.org/10.1103/PhysRevD.109.104026} {\bibfield  {journal} {\bibinfo
   {journal} {Phys. Rev. D}\ }\textbf {\bibinfo {volume} {109}},\ \bibinfo
  {pages} {104026} (\bibinfo {year} {2024})},\ \Eprint
  {https://arxiv.org/abs/2310.06033} {arXiv:2310.06033 [gr-qc]} \BibitemShut
  {NoStop}%
\bibitem [{Note1()}]{Note1}%
  \BibitemOpen
  \bibinfo {note} {In the metric approach, the perturbations are governed by
  the Regge-Wheeler \cite {regge1957stability} and Zerilli \cite
  {zerilli1970effective} equations, for axial and polar perturbations
  respectively. Although these equations share the same QNM spectrum in the
  case of Schwarschild black holes, isospectrality is not manifest because the
  equations take a very different form.}\BibitemShut {Stop}%
\bibitem [{\citenamefont {Teukolsky}(1972)}]{Teukolsky:1972my}%
  \BibitemOpen
  \bibfield  {author} {\bibinfo {author} {\bibfnamefont {S.~A.}\ \bibnamefont
  {Teukolsky}},\ }\bibfield  {title} {\bibinfo {title} {{Rotating black holes -
  separable wave equations for gravitational and electromagnetic
  perturbations}},\ }\href {https://doi.org/10.1103/PhysRevLett.29.1114}
  {\bibfield  {journal} {\bibinfo  {journal} {Phys. Rev. Lett.}\ }\textbf
  {\bibinfo {volume} {29}},\ \bibinfo {pages} {1114} (\bibinfo {year}
  {1972})}\BibitemShut {NoStop}%
\bibitem [{Note2()}]{Note2}%
  \BibitemOpen
  \bibinfo {note} {In fact, each of the equations can be transformed into the
  other by using a change of variables: the Teukolsky-Starobinsky identities
  \cite {Fiziev:2009ud}.}\BibitemShut {Stop}%
\bibitem [{\citenamefont {Chandrasekhar}(1984)}]{Chandrasekhar:1984siy}%
  \BibitemOpen
  \bibfield  {author} {\bibinfo {author} {\bibfnamefont {S.}~\bibnamefont
  {Chandrasekhar}},\ }\bibfield  {title} {\bibinfo {title} {{The Mathematical
  Theory of Black Holes}},\ }\href
  {https://doi.org/10.1007/978-94-009-6469-3_2} {\bibfield  {journal} {\bibinfo
   {journal} {Fundam. Theor. Phys.}\ }\textbf {\bibinfo {volume} {9}},\
  \bibinfo {pages} {5} (\bibinfo {year} {1984})}\BibitemShut {NoStop}%
\bibitem [{\citenamefont {Li}\ \emph {et~al.}(2023)\citenamefont {Li},
  \citenamefont {Wagle}, \citenamefont {Chen},\ and\ \citenamefont
  {Yunes}}]{Li:2022pcy}%
  \BibitemOpen
  \bibfield  {author} {\bibinfo {author} {\bibfnamefont {D.}~\bibnamefont
  {Li}}, \bibinfo {author} {\bibfnamefont {P.}~\bibnamefont {Wagle}}, \bibinfo
  {author} {\bibfnamefont {Y.}~\bibnamefont {Chen}},\ and\ \bibinfo {author}
  {\bibfnamefont {N.}~\bibnamefont {Yunes}},\ }\bibfield  {title} {\bibinfo
  {title} {{Perturbations of Spinning Black Holes beyond General Relativity:
  Modified Teukolsky Equation}},\ }\href
  {https://doi.org/10.1103/PhysRevX.13.021029} {\bibfield  {journal} {\bibinfo
  {journal} {Phys. Rev. X}\ }\textbf {\bibinfo {volume} {13}},\ \bibinfo
  {pages} {021029} (\bibinfo {year} {2023})},\ \Eprint
  {https://arxiv.org/abs/2206.10652} {arXiv:2206.10652 [gr-qc]} \BibitemShut
  {NoStop}%
\bibitem [{\citenamefont {Hussain}\ and\ \citenamefont
  {Zimmerman}(2022)}]{Hussain:2022ins}%
  \BibitemOpen
  \bibfield  {author} {\bibinfo {author} {\bibfnamefont {A.}~\bibnamefont
  {Hussain}}\ and\ \bibinfo {author} {\bibfnamefont {A.}~\bibnamefont
  {Zimmerman}},\ }\bibfield  {title} {\bibinfo {title} {{Approach to computing
  spectral shifts for black holes beyond Kerr}},\ }\href
  {https://doi.org/10.1103/PhysRevD.106.104018} {\bibfield  {journal} {\bibinfo
   {journal} {Phys. Rev. D}\ }\textbf {\bibinfo {volume} {106}},\ \bibinfo
  {pages} {104018} (\bibinfo {year} {2022})},\ \Eprint
  {https://arxiv.org/abs/2206.10653} {arXiv:2206.10653 [gr-qc]} \BibitemShut
  {NoStop}%
\bibitem [{\citenamefont {Cano}\ \emph
  {et~al.}(2023{\natexlab{b}})\citenamefont {Cano}, \citenamefont {Fransen},
  \citenamefont {Hertog},\ and\ \citenamefont {Maenaut}}]{Cano:2023jbk}%
  \BibitemOpen
  \bibfield  {author} {\bibinfo {author} {\bibfnamefont {P.~A.}\ \bibnamefont
  {Cano}}, \bibinfo {author} {\bibfnamefont {K.}~\bibnamefont {Fransen}},
  \bibinfo {author} {\bibfnamefont {T.}~\bibnamefont {Hertog}},\ and\ \bibinfo
  {author} {\bibfnamefont {S.}~\bibnamefont {Maenaut}},\ }\bibfield  {title}
  {\bibinfo {title} {{Quasinormal modes of rotating black holes in
  higher-derivative gravity}},\ }\href
  {https://doi.org/10.1103/PhysRevD.108.124032} {\bibfield  {journal} {\bibinfo
   {journal} {Phys. Rev. D}\ }\textbf {\bibinfo {volume} {108}},\ \bibinfo
  {pages} {124032} (\bibinfo {year} {2023}{\natexlab{b}})},\ \Eprint
  {https://arxiv.org/abs/2307.07431} {arXiv:2307.07431 [gr-qc]} \BibitemShut
  {NoStop}%
\bibitem [{Note3()}]{Note3}%
  \BibitemOpen
  \bibinfo {note} {This limit has to be taken with care, as the form of $\delta
  V_{s}$ is ambiguous due to the possibility of redefining the radial variables
  $R_{s}$. For instance, we find that $\delta V_{s}$ can diverge with a very
  high power of $l$, but the power of $l$ can be reduced by introducing
  appropriate transformations of the radial variable. We have used this kind of
  transformations to obtain the expressions (\ref {cubicdV}) and (\ref
  {quarticdV}). We also note that the limit involves taking $l\rightarrow
  \infty $ and $\omega \rightarrow \infty $ with $\omega /l$
  fixed.}\BibitemShut {Stop}%
\bibitem [{\citenamefont {Cano}\ and\ \citenamefont
  {David}(2024)}]{Cano:2024bhh}%
  \BibitemOpen
  \bibfield  {author} {\bibinfo {author} {\bibfnamefont {P.~A.}\ \bibnamefont
  {Cano}}\ and\ \bibinfo {author} {\bibfnamefont {M.}~\bibnamefont {David}},\
  }\bibfield  {title} {\bibinfo {title} {{Teukolsky equation for near-extremal
  black holes beyond general relativity: near-horizon analysis}},\ }\href@noop
  {} {\  (\bibinfo {year} {2024})},\ \Eprint {https://arxiv.org/abs/2407.02017}
  {arXiv:2407.02017 [gr-qc]} \BibitemShut {NoStop}%
\bibitem [{\citenamefont {Cardoso}\ \emph {et~al.}(2009)\citenamefont
  {Cardoso}, \citenamefont {Miranda}, \citenamefont {Berti}, \citenamefont
  {Witek},\ and\ \citenamefont {Zanchin}}]{Cardoso:2008bp}%
  \BibitemOpen
  \bibfield  {author} {\bibinfo {author} {\bibfnamefont {V.}~\bibnamefont
  {Cardoso}}, \bibinfo {author} {\bibfnamefont {A.~S.}\ \bibnamefont
  {Miranda}}, \bibinfo {author} {\bibfnamefont {E.}~\bibnamefont {Berti}},
  \bibinfo {author} {\bibfnamefont {H.}~\bibnamefont {Witek}},\ and\ \bibinfo
  {author} {\bibfnamefont {V.~T.}\ \bibnamefont {Zanchin}},\ }\bibfield
  {title} {\bibinfo {title} {{Geodesic stability, Lyapunov exponents and
  quasinormal modes}},\ }\href {https://doi.org/10.1103/PhysRevD.79.064016}
  {\bibfield  {journal} {\bibinfo  {journal} {Phys. Rev. D}\ }\textbf {\bibinfo
  {volume} {79}},\ \bibinfo {pages} {064016} (\bibinfo {year} {2009})},\
  \Eprint {https://arxiv.org/abs/0812.1806} {arXiv:0812.1806 [hep-th]}
  \BibitemShut {NoStop}%
\bibitem [{\citenamefont {Yang}\ \emph {et~al.}(2012)\citenamefont {Yang},
  \citenamefont {Nichols}, \citenamefont {Zhang}, \citenamefont {Zimmerman},
  \citenamefont {Zhang},\ and\ \citenamefont {Chen}}]{Yang:2012he}%
  \BibitemOpen
  \bibfield  {author} {\bibinfo {author} {\bibfnamefont {H.}~\bibnamefont
  {Yang}}, \bibinfo {author} {\bibfnamefont {D.~A.}\ \bibnamefont {Nichols}},
  \bibinfo {author} {\bibfnamefont {F.}~\bibnamefont {Zhang}}, \bibinfo
  {author} {\bibfnamefont {A.}~\bibnamefont {Zimmerman}}, \bibinfo {author}
  {\bibfnamefont {Z.}~\bibnamefont {Zhang}},\ and\ \bibinfo {author}
  {\bibfnamefont {Y.}~\bibnamefont {Chen}},\ }\bibfield  {title} {\bibinfo
  {title} {{Quasinormal-mode spectrum of Kerr black holes and its geometric
  interpretation}},\ }\href {https://doi.org/10.1103/PhysRevD.86.104006}
  {\bibfield  {journal} {\bibinfo  {journal} {Phys. Rev. D}\ }\textbf {\bibinfo
  {volume} {86}},\ \bibinfo {pages} {104006} (\bibinfo {year} {2012})},\
  \Eprint {https://arxiv.org/abs/1207.4253} {arXiv:1207.4253 [gr-qc]}
  \BibitemShut {NoStop}%
\bibitem [{\citenamefont {Bryant}\ \emph {et~al.}(2021)\citenamefont {Bryant},
  \citenamefont {Silva}, \citenamefont {Yagi},\ and\ \citenamefont
  {Glampedakis}}]{Bryant:2021xdh}%
  \BibitemOpen
  \bibfield  {author} {\bibinfo {author} {\bibfnamefont {A.}~\bibnamefont
  {Bryant}}, \bibinfo {author} {\bibfnamefont {H.~O.}\ \bibnamefont {Silva}},
  \bibinfo {author} {\bibfnamefont {K.}~\bibnamefont {Yagi}},\ and\ \bibinfo
  {author} {\bibfnamefont {K.}~\bibnamefont {Glampedakis}},\ }\bibfield
  {title} {\bibinfo {title} {{Eikonal quasinormal modes of black holes beyond
  general relativity. III. Scalar Gauss-Bonnet gravity}},\ }\href
  {https://doi.org/10.1103/PhysRevD.104.044051} {\bibfield  {journal} {\bibinfo
   {journal} {Phys. Rev. D}\ }\textbf {\bibinfo {volume} {104}},\ \bibinfo
  {pages} {044051} (\bibinfo {year} {2021})},\ \Eprint
  {https://arxiv.org/abs/2106.09657} {arXiv:2106.09657 [gr-qc]} \BibitemShut
  {NoStop}%
\bibitem [{Note4()}]{Note4}%
  \BibitemOpen
  \bibinfo {note} {The QNMs in the theory (\ref {eq:Lspecial}) are in fact only
  isospectral in the eikonal limit, as isospectrality is broken for lower $l$
  modes. Nevertheless, by using the numerical values of QNMs available in the
  literature \cite {Cano:2023jbk}, one can see that the isospectrality breaking
  in the theory (\ref {eq:Lspecial}) is pretty mild even for the lower $l$
  modes.}\BibitemShut {Stop}%
\bibitem [{\citenamefont {Grisaru}\ and\ \citenamefont
  {Zanon}(1986)}]{Grisaru:1986vi}%
  \BibitemOpen
  \bibfield  {author} {\bibinfo {author} {\bibfnamefont {M.~T.}\ \bibnamefont
  {Grisaru}}\ and\ \bibinfo {author} {\bibfnamefont {D.}~\bibnamefont
  {Zanon}},\ }\bibfield  {title} {\bibinfo {title} {{$\sigma$ Model Superstring
  Corrections to the Einstein-hilbert Action}},\ }\href
  {https://doi.org/10.1016/0370-2693(86)90765-3} {\bibfield  {journal}
  {\bibinfo  {journal} {Phys. Lett. B}\ }\textbf {\bibinfo {volume} {177}},\
  \bibinfo {pages} {347} (\bibinfo {year} {1986})}\BibitemShut {NoStop}%
\bibitem [{\citenamefont {Cano}\ \emph
  {et~al.}(2022{\natexlab{b}})\citenamefont {Cano}, \citenamefont {Ganchev},
  \citenamefont {Mayerson},\ and\ \citenamefont {Ruip\'erez}}]{Cano:2022wwo}%
  \BibitemOpen
  \bibfield  {author} {\bibinfo {author} {\bibfnamefont {P.~A.}\ \bibnamefont
  {Cano}}, \bibinfo {author} {\bibfnamefont {B.}~\bibnamefont {Ganchev}},
  \bibinfo {author} {\bibfnamefont {D.~R.}\ \bibnamefont {Mayerson}},\ and\
  \bibinfo {author} {\bibfnamefont {A.}~\bibnamefont {Ruip\'erez}},\ }\bibfield
   {title} {\bibinfo {title} {{Black hole multipoles in higher-derivative
  gravity}},\ }\href {https://doi.org/10.1007/JHEP12(2022)120} {\bibfield
  {journal} {\bibinfo  {journal} {JHEP}\ }\textbf {\bibinfo {volume} {12}},\
  \bibinfo {pages} {120}},\ \Eprint {https://arxiv.org/abs/2208.01044}
  {arXiv:2208.01044 [gr-qc]} \BibitemShut {NoStop}%
\bibitem [{Note5()}]{Note5}%
  \BibitemOpen
  \bibinfo {note} {Although the string theory effective action contains other
  fields, they can be consistently truncated at this order in $\alpha '$. In
  particular, the dilaton gets an $\protect \mathcal {O}(\alpha '^3)$ value due
  to its coupling to the $R^4$ term, but it only affects the metric at order
  $\protect \mathcal {O}(\alpha '^6)$.}\BibitemShut {Stop}%
\bibitem [{\citenamefont {Henneaux}\ and\ \citenamefont
  {Teitelboim}(2005)}]{Henneaux:2004jw}%
  \BibitemOpen
  \bibfield  {author} {\bibinfo {author} {\bibfnamefont {M.}~\bibnamefont
  {Henneaux}}\ and\ \bibinfo {author} {\bibfnamefont {C.}~\bibnamefont
  {Teitelboim}},\ }\bibfield  {title} {\bibinfo {title} {{Duality in linearized
  gravity}},\ }\href {https://doi.org/10.1103/PhysRevD.71.024018} {\bibfield
  {journal} {\bibinfo  {journal} {Phys. Rev. D}\ }\textbf {\bibinfo {volume}
  {71}},\ \bibinfo {pages} {024018} (\bibinfo {year} {2005})},\ \Eprint
  {https://arxiv.org/abs/gr-qc/0408101} {arXiv:gr-qc/0408101} \BibitemShut
  {NoStop}%
\bibitem [{\citenamefont {Hull}(2000)}]{Hull:2000rr}%
  \BibitemOpen
  \bibfield  {author} {\bibinfo {author} {\bibfnamefont {C.~M.}\ \bibnamefont
  {Hull}},\ }\bibfield  {title} {\bibinfo {title} {{Symmetries and
  compactifications of (4,0) conformal gravity}},\ }\href
  {https://doi.org/10.1088/1126-6708/2000/12/007} {\bibfield  {journal}
  {\bibinfo  {journal} {JHEP}\ }\textbf {\bibinfo {volume} {12}},\ \bibinfo
  {pages} {007}},\ \Eprint {https://arxiv.org/abs/hep-th/0011215}
  {arXiv:hep-th/0011215} \BibitemShut {NoStop}%
\bibitem [{\citenamefont {Born}\ and\ \citenamefont
  {Infeld}(1934)}]{Born:1934gh}%
  \BibitemOpen
  \bibfield  {author} {\bibinfo {author} {\bibfnamefont {M.}~\bibnamefont
  {Born}}\ and\ \bibinfo {author} {\bibfnamefont {L.}~\bibnamefont {Infeld}},\
  }\bibfield  {title} {\bibinfo {title} {{Foundations of the new field
  theory}},\ }\href {https://doi.org/10.1098/rspa.1934.0059} {\bibfield
  {journal} {\bibinfo  {journal} {Proc. Roy. Soc. Lond. A}\ }\textbf {\bibinfo
  {volume} {144}},\ \bibinfo {pages} {425} (\bibinfo {year}
  {1934})}\BibitemShut {NoStop}%
\bibitem [{\citenamefont
  {Bialynicki-Birula}(1984)}]{Bialynicki-Birula:1984daz}%
  \BibitemOpen
  \bibfield  {author} {\bibinfo {author} {\bibfnamefont {I.}~\bibnamefont
  {Bialynicki-Birula}},\ }\href@noop {} {\emph {\bibinfo {title}
  {{\emph{Nonlinear Electrodynamics: variations on a theme by Born and Infeld,
  in:} Quantum Theory of Particles and Fields: Birthday Volume Dedicated to Jan
  Lopuszanski}}}}\ (\bibinfo {year} {1984})\BibitemShut {NoStop}%
\bibitem [{\citenamefont {Gibbons}\ and\ \citenamefont
  {Rasheed}(1995)}]{Gibbons:1995cv}%
  \BibitemOpen
  \bibfield  {author} {\bibinfo {author} {\bibfnamefont {G.~W.}\ \bibnamefont
  {Gibbons}}\ and\ \bibinfo {author} {\bibfnamefont {D.~A.}\ \bibnamefont
  {Rasheed}},\ }\bibfield  {title} {\bibinfo {title} {{Electric - magnetic
  duality rotations in nonlinear electrodynamics}},\ }\href
  {https://doi.org/10.1016/0550-3213(95)00409-L} {\bibfield  {journal}
  {\bibinfo  {journal} {Nucl. Phys. B}\ }\textbf {\bibinfo {volume} {454}},\
  \bibinfo {pages} {185} (\bibinfo {year} {1995})},\ \Eprint
  {https://arxiv.org/abs/hep-th/9506035} {arXiv:hep-th/9506035} \BibitemShut
  {NoStop}%
\bibitem [{\citenamefont {Russo}\ and\ \citenamefont
  {Townsend}(2023)}]{Russo:2022qvz}%
  \BibitemOpen
  \bibfield  {author} {\bibinfo {author} {\bibfnamefont {J.~G.}\ \bibnamefont
  {Russo}}\ and\ \bibinfo {author} {\bibfnamefont {P.~K.}\ \bibnamefont
  {Townsend}},\ }\bibfield  {title} {\bibinfo {title} {{Nonlinear
  electrodynamics without birefringence}},\ }\href
  {https://doi.org/10.1007/JHEP01(2023)039} {\bibfield  {journal} {\bibinfo
  {journal} {JHEP}\ }\textbf {\bibinfo {volume} {01}},\ \bibinfo {pages}
  {039}},\ \Eprint {https://arxiv.org/abs/2211.10689} {arXiv:2211.10689
  [hep-th]} \BibitemShut {NoStop}%
\bibitem [{\citenamefont {Marques}\ and\ \citenamefont
  {Nunez}(2015)}]{Marques:2015vua}%
  \BibitemOpen
  \bibfield  {author} {\bibinfo {author} {\bibfnamefont {D.}~\bibnamefont
  {Marques}}\ and\ \bibinfo {author} {\bibfnamefont {C.~A.}\ \bibnamefont
  {Nunez}},\ }\bibfield  {title} {\bibinfo {title} {{T-duality and
  \ensuremath{\alpha'}-corrections}},\ }\href
  {https://doi.org/10.1007/JHEP10(2015)084} {\bibfield  {journal} {\bibinfo
  {journal} {JHEP}\ }\textbf {\bibinfo {volume} {10}},\ \bibinfo {pages}
  {084}},\ \Eprint {https://arxiv.org/abs/1507.00652} {arXiv:1507.00652
  [hep-th]} \BibitemShut {NoStop}%
\bibitem [{\citenamefont {Baron}\ \emph {et~al.}(2017)\citenamefont {Baron},
  \citenamefont {Fernandez-Melgarejo}, \citenamefont {Marques},\ and\
  \citenamefont {Nunez}}]{Baron:2017dvb}%
  \BibitemOpen
  \bibfield  {author} {\bibinfo {author} {\bibfnamefont {W.~H.}\ \bibnamefont
  {Baron}}, \bibinfo {author} {\bibfnamefont {J.~J.}\ \bibnamefont
  {Fernandez-Melgarejo}}, \bibinfo {author} {\bibfnamefont {D.}~\bibnamefont
  {Marques}},\ and\ \bibinfo {author} {\bibfnamefont {C.}~\bibnamefont
  {Nunez}},\ }\bibfield  {title} {\bibinfo {title} {{The Odd story of
  \ensuremath{\alpha'}-corrections}},\ }\href
  {https://doi.org/10.1007/JHEP04(2017)078} {\bibfield  {journal} {\bibinfo
  {journal} {JHEP}\ }\textbf {\bibinfo {volume} {04}},\ \bibinfo {pages}
  {078}},\ \Eprint {https://arxiv.org/abs/1702.05489} {arXiv:1702.05489
  [hep-th]} \BibitemShut {NoStop}%
\bibitem [{\citenamefont {Hohm}\ and\ \citenamefont
  {Zwiebach}(2016)}]{Hohm:2015doa}%
  \BibitemOpen
  \bibfield  {author} {\bibinfo {author} {\bibfnamefont {O.}~\bibnamefont
  {Hohm}}\ and\ \bibinfo {author} {\bibfnamefont {B.}~\bibnamefont
  {Zwiebach}},\ }\bibfield  {title} {\bibinfo {title} {{T-duality Constraints
  on Higher Derivatives Revisited}},\ }\href
  {https://doi.org/10.1007/JHEP04(2016)101} {\bibfield  {journal} {\bibinfo
  {journal} {JHEP}\ }\textbf {\bibinfo {volume} {04}},\ \bibinfo {pages}
  {101}},\ \Eprint {https://arxiv.org/abs/1510.00005} {arXiv:1510.00005
  [hep-th]} \BibitemShut {NoStop}%
\bibitem [{\citenamefont {Codina}\ \emph {et~al.}(2021)\citenamefont {Codina},
  \citenamefont {Hohm},\ and\ \citenamefont {Marques}}]{Codina:2020kvj}%
  \BibitemOpen
  \bibfield  {author} {\bibinfo {author} {\bibfnamefont {T.}~\bibnamefont
  {Codina}}, \bibinfo {author} {\bibfnamefont {O.}~\bibnamefont {Hohm}},\ and\
  \bibinfo {author} {\bibfnamefont {D.}~\bibnamefont {Marques}},\ }\bibfield
  {title} {\bibinfo {title} {{String Dualities at Order $\alpha'^{\,3}$}},\
  }\href {https://doi.org/10.1103/PhysRevLett.126.171602} {\bibfield  {journal}
  {\bibinfo  {journal} {Phys. Rev. Lett.}\ }\textbf {\bibinfo {volume} {126}},\
  \bibinfo {pages} {171602} (\bibinfo {year} {2021})},\ \Eprint
  {https://arxiv.org/abs/2012.15677} {arXiv:2012.15677 [hep-th]} \BibitemShut
  {NoStop}%
\bibitem [{\citenamefont {Garousi}(2021)}]{Garousi:2020gio}%
  \BibitemOpen
  \bibfield  {author} {\bibinfo {author} {\bibfnamefont {M.~R.}\ \bibnamefont
  {Garousi}},\ }\bibfield  {title} {\bibinfo {title} {{Effective action of type
  II superstring theories at order $\alpha'^{3}$: NS-NS couplings}},\ }\href
  {https://doi.org/10.1007/JHEP02(2021)157} {\bibfield  {journal} {\bibinfo
  {journal} {JHEP}\ }\textbf {\bibinfo {volume} {02}},\ \bibinfo {pages}
  {157}},\ \Eprint {https://arxiv.org/abs/2011.02753} {arXiv:2011.02753
  [hep-th]} \BibitemShut {NoStop}%
\bibitem [{\citenamefont {David}\ and\ \citenamefont
  {Liu}(2022)}]{David:2021jqn}%
  \BibitemOpen
  \bibfield  {author} {\bibinfo {author} {\bibfnamefont {M.}~\bibnamefont
  {David}}\ and\ \bibinfo {author} {\bibfnamefont {J.~T.}\ \bibnamefont
  {Liu}},\ }\bibfield  {title} {\bibinfo {title} {{T duality and hints of
  generalized geometry in string \ensuremath{\alpha}' corrections}},\ }\href
  {https://doi.org/10.1103/PhysRevD.106.106008} {\bibfield  {journal} {\bibinfo
   {journal} {Phys. Rev. D}\ }\textbf {\bibinfo {volume} {106}},\ \bibinfo
  {pages} {106008} (\bibinfo {year} {2022})},\ \Eprint
  {https://arxiv.org/abs/2108.04370} {arXiv:2108.04370 [hep-th]} \BibitemShut
  {NoStop}%
\bibitem [{\citenamefont {David}\ and\ \citenamefont
  {Liu}(2023)}]{David:2022jcl}%
  \BibitemOpen
  \bibfield  {author} {\bibinfo {author} {\bibfnamefont {M.}~\bibnamefont
  {David}}\ and\ \bibinfo {author} {\bibfnamefont {J.~T.}\ \bibnamefont
  {Liu}},\ }\bibfield  {title} {\bibinfo {title} {{T-duality building blocks
  for \ensuremath{\alpha}' string corrections}},\ }\href
  {https://doi.org/10.1103/PhysRevD.107.046008} {\bibfield  {journal} {\bibinfo
   {journal} {Phys. Rev. D}\ }\textbf {\bibinfo {volume} {107}},\ \bibinfo
  {pages} {046008} (\bibinfo {year} {2023})},\ \Eprint
  {https://arxiv.org/abs/2210.16593} {arXiv:2210.16593 [hep-th]} \BibitemShut
  {NoStop}%
\bibitem [{\citenamefont {Mart\'in-Garc\'ia}()}]{xAct}%
  \BibitemOpen
  \bibfield  {author} {\bibinfo {author} {\bibfnamefont {J.~M.}\ \bibnamefont
  {Mart\'in-Garc\'ia}},\ }\href {http://www.xact.es/} {\bibinfo {title} {{xAct:
  Efficient tensor computer algebra for the Wolfram Language}}},\ \bibinfo
  {howpublished} {\url{http://www.xact.es/}}\BibitemShut {NoStop}%
\bibitem [{\citenamefont {Reall}(2021)}]{Reall:2021voz}%
  \BibitemOpen
  \bibfield  {author} {\bibinfo {author} {\bibfnamefont {H.~S.}\ \bibnamefont
  {Reall}},\ }\bibfield  {title} {\bibinfo {title} {{Causality in gravitational
  theories with second order equations of motion}},\ }\href
  {https://doi.org/10.1103/PhysRevD.103.084027} {\bibfield  {journal} {\bibinfo
   {journal} {Phys. Rev. D}\ }\textbf {\bibinfo {volume} {103}},\ \bibinfo
  {pages} {084027} (\bibinfo {year} {2021})},\ \Eprint
  {https://arxiv.org/abs/2101.11623} {arXiv:2101.11623 [gr-qc]} \BibitemShut
  {NoStop}%
\bibitem [{\citenamefont {Regge}\ and\ \citenamefont
  {Wheeler}(1957)}]{regge1957stability}%
  \BibitemOpen
  \bibfield  {author} {\bibinfo {author} {\bibfnamefont {T.}~\bibnamefont
  {Regge}}\ and\ \bibinfo {author} {\bibfnamefont {J.~A.}\ \bibnamefont
  {Wheeler}},\ }\bibfield  {title} {\bibinfo {title} {Stability of a
  schwarzschild singularity},\ }\href@noop {} {\bibfield  {journal} {\bibinfo
  {journal} {Physical Review}\ }\textbf {\bibinfo {volume} {108}},\ \bibinfo
  {pages} {1063} (\bibinfo {year} {1957})}\BibitemShut {NoStop}%
\bibitem [{\citenamefont {Zerilli}(1970)}]{zerilli1970effective}%
  \BibitemOpen
  \bibfield  {author} {\bibinfo {author} {\bibfnamefont {F.~J.}\ \bibnamefont
  {Zerilli}},\ }\bibfield  {title} {\bibinfo {title} {Effective potential for
  even-parity regge-wheeler gravitational perturbation equations},\ }\href@noop
  {} {\bibfield  {journal} {\bibinfo  {journal} {Physical Review Letters}\
  }\textbf {\bibinfo {volume} {24}},\ \bibinfo {pages} {737} (\bibinfo {year}
  {1970})}\BibitemShut {NoStop}%
\bibitem [{\citenamefont {Fiziev}(2009)}]{Fiziev:2009ud}%
  \BibitemOpen
  \bibfield  {author} {\bibinfo {author} {\bibfnamefont {P.~P.}\ \bibnamefont
  {Fiziev}},\ }\bibfield  {title} {\bibinfo {title} {{Teukolsky-Starobinsky
  Identities: A Novel Derivation and Generalizations}},\ }\href
  {https://doi.org/10.1103/PhysRevD.80.124001} {\bibfield  {journal} {\bibinfo
  {journal} {Phys. Rev. D}\ }\textbf {\bibinfo {volume} {80}},\ \bibinfo
  {pages} {124001} (\bibinfo {year} {2009})},\ \Eprint
  {https://arxiv.org/abs/0906.5108} {arXiv:0906.5108 [gr-qc]} \BibitemShut
  {NoStop}%
\end{thebibliography}%
\noindent \centering

\end{document}
%